\documentclass{aa}  

\usepackage{graphicx}
\graphicspath{ {./figure/} }
\usepackage{txfonts}
\usepackage[dvipsnames]{xcolor}
\usepackage{hyperref}
\usepackage{algorithm}
\usepackage{algpseudocode}
\usepackage{float}
\usepackage{ulem}
\usepackage{rotating}
\usepackage{subcaption}
\usepackage{mwe}
\usepackage{listings}
\usepackage{booktabs}
\usepackage{pgf}
\usepackage{soul}
\setstcolor{red}

\algnewcommand\algorithmicforeach{\textbf{for each}}
\algdef{S}[FOR]{ForEach}[1]{\algorithmicforeach\ #1\ \algorithmicdo}

\newcommand{\review}[1]{{\textcolor{black}{#1}}}

\newcommand{\Traj}{{\cal{Q}}}
\newcommand{\Rel}{{\cal{R}}}
\newcommand{\Alert}{{\cal{A}}}
\newcommand{\OldAlert}{{\cal{O}}}
\newcommand{\traj}{q}
\newcommand{\orbfit}{\texttt{OrbFit}}

\definecolor{codegreen}{rgb}{0,0.6,0}
\definecolor{codegray}{rgb}{0.5,0.5,0.5}
\definecolor{codepurple}{rgb}{0.58,0,0.82}
\definecolor{backcolour}{rgb}{0.95,0.95,0.92}
\lstdefinestyle{mystyle}{
    commentstyle=\color{codegray},
    keywordstyle=\color{magenta},
    numberstyle=\tiny\color{codegray},
    stringstyle=\color{codegreen},
    basicstyle=\ttfamily\footnotesize,
    breakatwhitespace=false,         
    breaklines=true,                 
    captionpos=b,                    
    keepspaces=true,                 
    showspaces=false,                
    showstringspaces=false,
    showtabs=false,                  
    tabsize=2
}
\begin{document}

   \title{Enabling discoveries of Solar System objects in large alert data streams}

   \author{
        R. Le Montagner\inst{1}\fnmsep\thanks{\email{roman.le-montagner@ijclab.in2p3.fr}}
    \and
        J. Peloton\inst{1}
    \and
        B. Carry\inst{2}
    \and 
        J. Desmars\inst{3,4}
    \and
        D. Hestroffer\inst{4}
    \and
        R. A. Mendez\inst{5}
    \and
        A. C. Perlbarg\inst{3,4}
    \and
        W. Thuillot\inst{4}
    }

   \institute{
        Universit\'e Paris-Saclay, CNRS/IN2P3, IJCLab, 91405 Orsay, France
    \and
        Universit{\'e} C{\^o}te d'Azur, Observatoire de la C{\^o}te d'Azur, CNRS, Laboratoire Lagrange, France
    \and
        Institut Polytechnique des Sciences Avanc\'ees IPSA, 63 boulevard de Brandebourg, F-94200 Ivry-sur-Seine, France
    \and
        IMCCE, Observatoire de Paris, Universit\'e PSL, CNRS, Sorbonne Universit\'e, Univ. Lille, 77 av. Denfert-Rochereau, 75014 Paris, France
    \and
        Departamento de Astronomía, Universidad de Chile, Casilla 36-D, Correo Central, Santiago, Chile 
    }

   \date{Received September 15, 1996; accepted March 16, 1997}
 
  \abstract
   {
   \review{With the advent of large-scale astronomical surveys, such as the Zwicky Transient Facility (ZTF) and the forthcoming Vera C. Rubin Observatory's Legacy Survey of Space and Time (LSST)}, the number of alerts generated by transient, variable, and moving astronomical objects is growing rapidly, reaching millions of alerts per night. Concerning the minor planets of the Solar System, their identification requires linking the alerts for many observations over a potentially lengthy period of time, leading to a very large combinatorial number.
   }
   {
   \review{The goal is to demonstrate how a third-party module dedicated to the identification of new minor planets of the Solar System  can be integrated with the Fink alert broker real-time operations, which deals with massive alert data streams produced by large-scale surveys.}
   }
   {
   Our analysis takes advantage of the \review{scientific surplus brought on by the} Fink alert broker classification capabilities to first reduce the 111,275,131 processed alerts from ZTF between November 2019 and December 2022 (755 observation nights) to only 389,530 new Solar System alert candidates over the same period. We implemented a \review{simple, yet pedagogical} linking algorithm called Fink-FAT to create trajectory candidates in real time from alert data and extract orbital parameters. The analysis was validated on ZTF alert packets linked to confirmed Solar System objects from the Minor Planet Center (MPC) database. Finally, the \review{candidates} were confronted with follow-up observations.
   }
   {
   Between November 2019 and December 2022, Fink-FAT extracted 327 new orbits from candidate Solar System objects at the time of the observations, of which 65 had still remained unreported in the MPC database as of March 2023. After two late follow-up observation campaigns of six orbit candidates, four were associated with known minor planets of the Solar System, and two still remain unknown. In terms of performance, Fink-FAT took under 3h to link alerts into trajectory candidates and to extract the orbital elements over the three years of Fink data, using a modest hardware configuration.
   }
   {
   \review{Despite a much lower efficiency than present linking algorithms, Fink-FAT reaches a high level of purity in reconstructing orbits and it runs fast, making it suitable for the real-time discovery of new minor planets. Fink-FAT is deployed in the Fink broker and analyzes, in real time, the alert data from the ZTF survey by regularly extracting new candidates for Solar System objects. Tests of scalability also show that Fink-FAT is capable of handling the even larger volume of alert data that will be sent by the Rubin Observatory's real-time difference image analysis processing.}
   }


   \maketitle
%

\section{Introduction}

Recent optical surveys such as the Zwicky Transient Facility (ZTF) \citep{2019PASP..131a8003M,2019PASP..131g8001G,2019PASP..131a8002B,2019PASP..131a8001P} and Pan-STARRS \citep{Denneau_2013} generate alerts by detecting differences from previous observations of the same areas of the sky. These alerts must be released early on to enable a rapid response from follow-up facilities when necessary; hence, they contain a minimal amount of information, namely: the observation time, sky coordinates, and estimation of the brightness.
Among its many applications, the analysis of these alerts by the scientific community enables the study of the Solar System's small bodies, which, in turn, allows for example a better understanding of the dynamical evolution of the Solar System
\citep{2014Natur.505..629D, 2015aste.book..493M}. Every night, new observations provide additional information to known Solar System objects or lead to the discovery of new objects. 

Naively, the identification of Solar System objects from difference imaging techniques requires linking the alerts of many observations over a potentially large period of time, leading to a very large combinatorial number. While we are already facing technical challenges due to large volumes of data, the exponential increase in the volume of data driven by upcoming large optical surveys such as Vera C. Rubin Observatory's Legacy Survey of Space and Time (LSST) \citep{2009arXiv0912.0201L, schwamb2023tuning} will strengthen the challenges and hinder the scientific exploitation of the data sets. To overcome the challenges posed by the linkage problems in the context of large volumes of alert data, several methods have been proposed over the last decade. For example, to make the problem more computationally feasible, survey cadence strategy can be adapted to systematically take observations of the same fields spaced by a predefined time window, depending on the targeted type of objects, and typically ranging from less than an hour for inner Solar System objects to more spaced cadence for outer objects (see e.g., \cite{2016AJ....152...70B}).
This design allows for the construction of so-called tracklets for moving objects when differencing the two observation images\footnote{A tracklet is a sequence of 2 or more spatially nearby detections taken over a short time span and likely to be related to the same moving object.}. These tracklets, which contain information on the direction and the rate of motion and which are less numerous than the initial number of alerts, are then linked to candidate orbits. This idea was first proposed and implemented in the Moving Object Processing System (MOPS), which produces automatic asteroid discoveries and identification for the Pan-STARRS survey \citep{2007Icar..189..151K, Denneau_2013}. However, despite the success of the method, it suffers many problems among which the number of orbit fits that must be carried out scales as $\mathcal{O}(N^3)$, where $N$ is the number of tracklets. For surveys producing millions of tracklets, this procedure becomes almost intractable. Since then, many alternatives have been proposed to improve the efficiency of the linking problem such as HelioLinC \citep{2018AJ....156..135H} and Heliolinc3D \cite{2022DPS....5450404H}. HelioLinC is a method that operates a change of the reference frame (topocentric to heliocentric) for linking detections, and propagates tracklets to common epochs to ease the identification of tracklets tracing the same underlying Solar System object's motion. In addition, HelioLinC reduces the complexity of the linking problem to $\mathcal{O}(N \log N)$, where $N$ is the number of tracklets, making it desirable in the context of large surveys. A modified version of HelioLinC has been successfully used in the context of HITS \citep{Pena_2018, Pena_2020}. However, similarly to MOPS, HelioLinC relies on the existence of tracklets, which put high constraints on the survey strategy design. Other methods relying on tracklets have been proposed such as CANFind \citep{2021AJ....162..244F}, using a technique directly based on the Hough Transform \citep{2020AJ....159...25L}. Another popular alternative to MOPS is the ZTF's Moving Object Discovery Engine (ZMODE) developed for the Palomar
Transient Factory (PTF) and scaled to meet the requirements of the ZTF survey \citep{2019PASP..131a8003M}. One of the main difference with MOPS is the construction of stringlets, which are a more flexible version of tracklets and better adapted to the cadence strategy of ZTF. More recently, the Tracklet-less Heliocentric Orbit Recovery (THOR) \citep{moeyens2021thor} algorithm proposed a solution inspired from HelioLinC, but without the need for intra-night linking (tracklets or stringlets). In addition, they operate a different change of the reference frame to linearize the motion of objects and use line-detection algorithm to identify orbits. Finally, other methods make use of specialized coprocessors such as graphics processing units (GPU) to accelerate the computation, such as the Kernel-Based Moving Object Detection (KBMOD) \citep{whidden2019fast} and its extension \citep{2021AJ....162..245S}.

\review{In this work, we do not attempt to find a new or better linking algorithm; rather, we describe how to easily extend an existing alert broker to enable third-party scientists to deploy and apply  a small body linking code on alert streams in real time. The use of a broker brings two major advantages: users can access alert data without having to obtain special access from the upstream surveys and the broker provides a scientific surplus used to provide an initial classification of alerts, hopefully redirecting only alerts of interest for new discoveries. These two leave more flexibility to the users for the identification of new minor planets in real-time.} We use the Fink broker\footnote{\url{https://fink-broker.org}}, whose original goal is to process large alert data streams, enrich them with information from other surveys and catalogs as well as machine-learning classification scores, and select the most promising events to follow for a wide-variety of science cases \citep{fink}. As opposed to traditional broker analysis techniques operating on commodity hardware, Fink implements a new technological approach by operating in real time on large computing infrastructures to enable a systemic analysis of the transient and variable sky from the Solar System objects to galactic and extra-galactic events. Since 2019, Fink has been analysing  the alert data stream from the ZTF optical time-domain survey in real time, and it is preparing to analyse the Rubin Observatory data stream in the coming years\footnote{\url{https://www.lsst.org/scientists/alert-brokers}}. \review{It is important to note though that other similar initiatives in this area exist, such as the SNAPS broker \cite{Trilling_2023} and the Asteroid Discovery Analysis and Mapping (ADAM\footnote{\url{https://adam.b612.ai/}}) platform. Yet one of the major advantages of Fink is the global study of the transient sky by coupling multiple data sources and simultaneously studying  various scientific areas, which brings the scientific surplus necessary to seamlessly classify the gigantic alert streams coming from deep and wide field surveys.}

The paper is organized as follows. In Section \ref{section_fink_fat}, we describe a simple yet efficient linking algorithm, called Fink-FAT, used to extract orbit candidate trajectories from alert data tagged as Solar System candidates and the fitting procedure used to compute the orbital parameters. \review{We also describe how Fink-FAT integrates within Fink}. Section \ref{section_sso} describes the alert data from Solar System objects collected by Fink from the ZTF alert stream. Section \ref{section_results} presents the performance of Fink-FAT on ZTF alert data, both in terms of computation time and recovery of known trajectories. Finally in Section \ref{section_discovery}, we present two follow-up campaigns focusing on previously unreported Solar System object candidates selected by Fink-FAT. 
 
\section{Fink-FAT: Fink Asteroid Tracker} \label{section_fink_fat}

Fink-FAT is a system dedicated to detect moving objects such as asteroids from a set of alerts emitted at different epochs. As a result, Fink-FAT returns a set of trajectories where alerts are linked based on a set of criteria. The system is also able to fit for an orbit based on these linked alerts. It is currently deployed and used within the Fink broker \citep{fink}. Each night, the system produces either new trajectories or continues the existing trajectories by adding new alerts. Fink-FAT also comes with an offline mode where the data from an arbitrary number of previous nights can be analysed together. In this section, we describe how the candidate trajectories are created in Fink-FAT from generic alert data, and the fitting procedure used to compute the orbital parameters.

\subsection{Alert association} \label{sec:assoc_steps}

Fink-FAT works in two phases (see Appendix \ref{app:pseudo_code} for the pseudo-code). The first phase is called the association and it forms a set of trajectories by linking all the alerts between them. The purpose of the association algorithm is not to find asteroids precisely but a set of coherent trajectories that behave like moving objects. To reduce the number of possible associations between alerts, the association algorithm relies on a set of three conditions (apparent motion, magnitude, and co-linearity) based on information from the incoming alerts such as: the position in equatorial coordinates (right ascension and declination), the apparent magnitude, the filter band identifier used during the exposure and the Julian date corresponding to the start exposure time. 

\subsubsection{Associating alerts} \label{sec:conditions}

First, the association of two alerts is done by spatial proximity. A KD-tree is used to efficiently perform the search of associations between thousands of alerts. All the alerts with a sky angular separation between them less than a specific threshold are associated. This search can generate many associations per alert. We let $\Delta d$ be the separation between two alerts separated in time by $\Delta t$, they are associated together by Fink-FAT if their separation satisfies the following condition:

 \begin{equation}
 \tag{condition 1}
     \dfrac{\Delta d}{\Delta t} < r_d,
 \end{equation}
where $r_d$ is a reference apparent motion rate (deg/day), and its value mainly depends on the targeted Solar System object population, and it is discussed in Section \ref{section_sso}.  

The second condition is based on the physical evolution of the asteroid luminosity. From observations, we can set boundaries on the expected change in magnitude between two observations of the same object. We let $\Delta m$ be the difference in magnitude between two alerts separated in time by $\Delta t$, we associate the two alerts if they satisfy the magnitude condition:

\begin{equation}
\tag{condition 2}
\bigg | \dfrac{\Delta m}{\Delta t} \bigg | < r_m ,
\end{equation}
where $r_m$ is a reference magnitude rate (mag/day) depending on the targeted population (see Sect. \ref{sec:sso_confirmed}). We note that the value of the rate also depends on the filter bands of each alert. In practice, this definition is only meaningful over a short period of time as the observed magnitude of objects oscillates because of their mostly non-spherical shape. The third condition is based on the dynamic of the object. The algorithm computes an angle $\alpha$ between the two last alert positions (in equatorial coordinates) of a potential trajectory and the new associated alerts separated by $\Delta t$ days, and the new alert is associated with the trajectory only if the following co-linearity condition is met: 

\begin{equation}
\tag{condition 3}
\dfrac{\alpha}{\Delta t} < r_\alpha.
\end{equation}
The choice for $r_\alpha$ (deg/day) is discussed in Section \ref{section_sso}, but we usually choose a small value (see, e.g., Table \ref{tab:sso_condition_inter}). Due to geometric projection, Solar System objects can produce complex trajectories in equatorial coordinates. However, over a small period of time (i.e., if frequent observations are performed), we suppose that the trajectories evolves smoothly, and the three conditions limit the number of false associations. 

\subsubsection{Starting a trajectory}

We let $\Traj$ be the set of all trajectories returned by Fink-FAT, and $\traj \in \Traj$ is a n-uplet of alerts linked together and supposedly coming from the same Solar System object. Fink-FAT starts a trajectory in two different ways. The first is the intra-night association step that defines a relation over the alerts coming from the same night. If the telescope observes repeatedly the same area on the sky (or adjacent areas), it allows us forming trajectories from the same observation night. 

We let $\Alert_i$ be the set of alerts coming from the night $i \in \mathbb{N}$, and $a_j \in \Alert_i$ an alert.
We define the intra-night relationship as:
\begin{equation}
\Rel_{intra}=\{(a_j, a_k) \, | \, \forall a_j, a_k \in \Alert_i, \, \text{condition 1} \wedge \text{condition 2} \}.
\end{equation}
The intra-night relation is reflexive, symmetric, and more important transitive, allowing the intra-night step to return trajectories larger than just pairs of points. Consequently, the intra-night association step returns a set of trajectories defined as:  
\begin{equation}
    \Traj_{intra}=\{\traj = (a_0, a_1, ... , a_k) \, | \, \forall a_k \in \Alert_i, a_k \Rel_{intra} a_{k+1} \}.
\end{equation}

The second way to start a trajectory is by associating alerts between different observation nights. Depending on the cadence of the telescope, and the motion of objects, there could be several days between two subsequent observations of the same object on the sky. We let $\OldAlert$ to be expressed as the set of old non-associated alerts:

\begin{equation}
   \OldAlert=\{a | a \in \bigcup\limits_{j=0}^{i - 1} \Alert_j \setminus \Traj \} 
.\end{equation}
The inter-night association define a new relation call $R_{inter}$:

\begin{equation}
\Rel_{inter} = \{ (a_i, a_j) | \forall a_i \in \OldAlert, \forall a_j \in \Alert_{i}, \text{condition 1} \wedge \text{condition 2} \}.
\end{equation}
The $\Rel_{inter}$ relation is also reflexive, symmetric, and transitive, but, unlike the $\Rel_{intra}$ relation, the $\Rel_{inter}$ relation does not use the transitivity and returns -- only pairs of alerts. Consequently, the inter-night association's step returns a set of pairs of points defined as:
\begin{equation}
\Traj_{inter}=\{(a_j, a_k) | \forall a_j \in \OldAlert, \forall a_k \in \Alert_{i}, a_j \Rel_{inter} a_{k} \}.
\end{equation}

\begin{figure}[h]
\includegraphics[width=9cm]{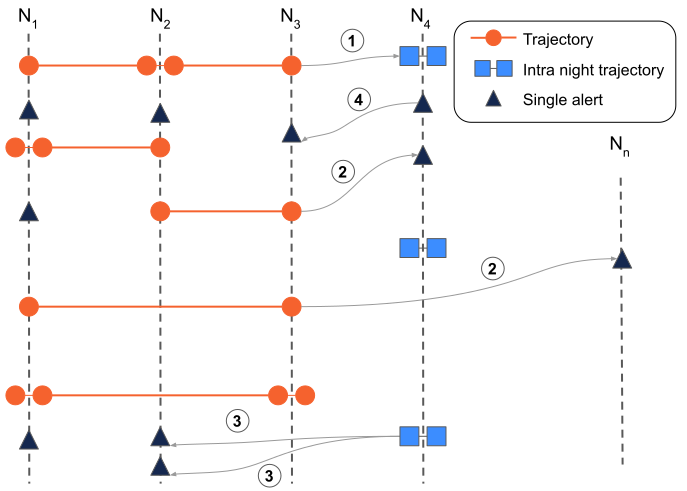}
\caption{Summary of the associations carried out by Fink-FAT. Each night is represented by a vertical dashed night denoted $N_i$. Alerts are represented by colored circles. The color-coding describes a type of association, shown in the legend of the plot.
The association step in Fink-FAT uses a sequential algorithm (1 $\rightarrow$ 2 $\rightarrow$ 3 $\rightarrow$ 4); therefore, the association order is important, especially since the previous step will remove the associated elements (trajectories, intra-night trajectories or single alert) from the possible association for the next steps.
The first step (1) is the association between the trajectories built from the previous night's alerts with the intra-night trajectories constructed during the current night. 
The second step (2) is the association between the trajectories and the remaining single alerts after the intra-night trajectories creation.
The third (3) and fourth (4) steps are similar, as they associate past alerts with current ones. The third step associates the intra-night trajectory's extremity with the old alerts. The fourth step associates current non-associated alerts with old alerts. The fourth step is one of the ways to start a trajectory as the intra-night trajectory building step. 
Note: each step can produce internally different trajectories including the same alert, as shown with the double association (3) at the bottom.
}
\label{fig:association_diagram}
\end{figure}

\subsubsection{Continuing a trajectory}

The next goal of Fink-FAT is to extend trajectories with alerts coming from new observations. There are three ways to continue an existing trajectory, as summarized in Fig. \ref{fig:association_diagram}. 

The first is the addition of a new intra-night trajectory to an existing trajectory. Two trajectories are merged by using their extremity. The addition is done using all conditions defined above. We let $\traj_i=(a_0, a_1, ..., a_k) \in \Traj$ be an existing trajectory, and $\traj_j=(b_0, b_1, ..., b_k) \in \Traj_{intra}$ a new intra-night trajectory. The new resulting trajectory is $\traj=(a_0, a_1, ..., a_k, b_0, b_1, ..., b_k),$ where $a_{k-1}, a_k, b_0$ satisfy the predicate $P(a_{k-1}, a_k, b_0), P=\text{condition 1} \wedge \text{condition 2} \wedge \text{condition 3}$.

The second way of continuing a trajectory is by adding a single alert to existing trajectories. As above, the addition of a new alert to an existing trajectory is done with the alert from the extremity of the existing trajectory. We let $\traj_i=(a_0, a_1, ..., a_k) \in \Traj$ be an existing trajectory and $b_i \in \Alert_i$ be an alert from the set of new incoming alerts. The resulting trajectory is $\traj=(a_0, a_1, ..., a_k, b_i),$ where $a_{k-1}, a_k, b_i$ satisfy the predicate $P(a_{k-1}, a_k, b_i)$.

Finally, the third and last way to continue a trajectory is by adding a single point to an intra-night trajectory. The purpose of this association is the same as above: adding a single point if the telescope does not come back twice to a field during the same night. Letting $t_i=(a_0, a_1, ..., a_k) \in T_{intra}$ and $b_i \in O$, $a_0, a_1, ..., a_k \in A_i$, the resulting trajectories are $t=(b_i, a_0, a_1, ..., a_k),$ where $b_i, a_0, a_1$ satisfy the predicate $P(b_i, a_0, a_1)$.

\subsubsection{Time window} \label{sec:time_window}

The formalism introduced above supposes to create trajectories by using all the alerts of the surveys, at all steps of the process. Despite the undeniable help brought by the broker system that will provide only relevant alerts to Fink-FAT by filtering out already classified alerts, the procedure above becomes computationally hard and inefficient for modern surveys such as the ZTF or the forthcoming LSST, as the number of possible associations each night grows exponentially. Therefore, Fink-FAT allows alert associations and keeps the trajectories in memory only during finite times (the impact is discussed in Sect. \ref{sec:time_impact}). In practice, we used three time window parameters: the separating time between the end of a trajectory and a new alert, the time to keep an old alert as candidate, and the time to keep an intra-night as candidate. 

\subsection{Orbit fitting} \label{sec:orbfit_description}

The second step of Fink-FAT is the orbit fitting. This step allows us to filter the trajectories that do not behave like asteroids from a physical point of view and it returns a set of orbital elements that describe the trajectory dynamics in the Solar System. Fink-FAT uses the \orbfit\ Software from The \orbfit\ Consortium \footnote{\href{OrbFit website}{\url{http://adams.dm.unipi.it/orbfit/}}}. 

Orbit determination is done in two steps. First, the initial orbit parameters are extracted using V\"ais\"al\"a's method to solve Gauss' problem of the orbit from three observations \citep{1985AJ.....90.1541M}. The method uses sets of three RA/Dec measurements and timings to determine an initial orbit, assuming a Keplerian motion. Once the parameters of the initial orbit have been estimated (if possible), a full differential correction step is performed to increase the accuracy of the initial computed orbital elements and estimate the covariance of the parameters. If the full differential corrections fail, we still retain the initial solution for short term predictions. In addition, we note that the public version of the software cannot compute the orbits of the satellites of planets.

\orbfit\ internally produces many files, and in the case of large number of observations to process, the read and write operations on internally generated files (I/O) take a significant part of the orbit fitting process. Choosing a RAM location can speed up the processing and preserve the lifetime of disks, making the orbit fitting essentially a CPU limited task. \orbfit\ takes 0.5 seconds on average on one modern core to fit one trajectory, that is it can process 1,000 trajectories with a modern eight-core laptop in about a minute with multiprocessing capabilities. While this is an acceptable rate regarding the data from current surveys, this will not be enough at the LSST era. Hence, Fink-FAT has also been extended to use \orbfit\ on clusters of machines to fit orbits of hundreds of trajectories simultaneously. This mode makes use of the framework Apache Spark \footnote{\url{https://spark.apache.org/}} to distribute the load and we made extensive tests on the VirtualData cloud of the Paris-Saclay University.

\begin{figure*}[h]
\centering
\includegraphics[width=0.3\textwidth]{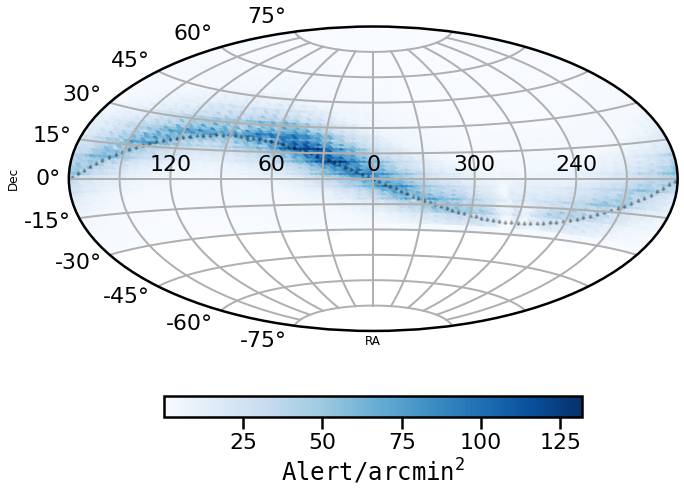}
\includegraphics[width=0.3\textwidth]{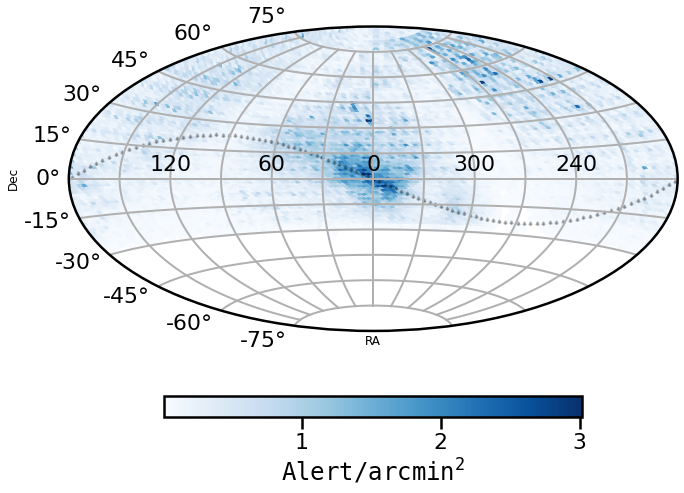}
\includegraphics[width=0.3\textwidth]{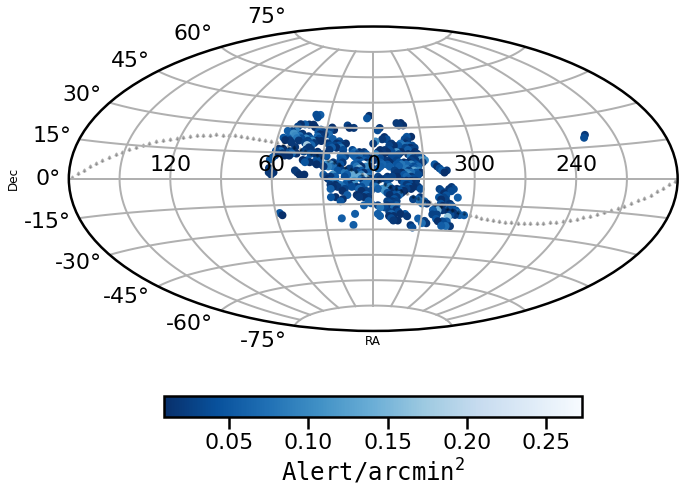}
\caption{Footprint of the ZTF alert stream from November 2019 to December 2022 associated with different subsets: the 15,381,246 alerts associated with confirmed Solar System objects (left; see Sect. \ref{sec:sso_confirmed}), the 389,530 alerts associated with Solar System object candidates (middle; see Sect. \ref{sec:sso_candidates}), and the 2,205 alerts associated with reconstructed orbits (right; see Sect. \ref{section_discovery}). The sky maps are in equatorial coordinates, and ZTF does not observe for declination lower than $\approx$ -30 degrees. For each footprint, we use the HEALPix pixelisation algorithm with a resolution parameter equals to \texttt{Nside=32} \citep{Gorski_2005} and the color scheme displays the number of alert per arcminute square. The color scale for the rightmost footprint has been inverted compared to the two others for a better readability. For reference, the ecliptic plane is shown with black triangles.}
\label{fig:skymap}
\end{figure*}

\subsection{Integration within the Fink ecosystem}

\review{Fink-FAT is an independent package from the main Fink code base\footnote{\url{https://github.com/FusRoman/fink-fat}}. It is installed as a dependency (version controlled) in the platform where Fink is running and it is called within the main schedule of the broker. At the end of each night, all alerts satisfying the Solar System candidate criteria are processed by Fink-FAT and results are automatically stored in the Fink database. Dedicated tables in the database are used for storing linked alerts and estimated orbital parameters and all results are available to Fink users via the different Fink services (science portal, REST API, data transfer service, and livestream service). The additional value created by Fink-FAT is also used by other scientific modules to improve their analysis, for inst the modules focusing on optical counterparts to gravitational wave events or gamma ray burst events are filtering out candidates selected by Fink-FAT.}

\subsection{Reasons for implementing another linking code}

\review{As we  explain in Section \ref{section_results}, Fink-FAT is not competitive in terms of reconstruction performances with respect to the present linking codes such as THOR \cite{moeyens2021thor} or Heliolinc3D \cite{2022DPS....5450404H}. Our goal is to have a code pedagogical and simple enough to focus on the integration with Fink operations, leaving the optimization of performance for a future work. Nevertheless Fink-FAT has the advantage of being open-source, it has a simple and intuitive implementation in Python which allows to appreciate the various challenges posed by the linking problem, the documentation is available online, and it runs fast with modest resources for our purpose. It ought to be noted that Fink-FAT is still a work in progress and improvements are foreseen (see Sect. \ref{sec:limitations}).}

\section{Solar system objects in Fink} \label{section_sso}

Each night, ZTF generates an unfiltered, 5 sigma alert stream extracted from difference images. Alerts are generated after each 30-second exposure and sent shortly after. They contain basic information such as the location of the transient on the sky or its magnitude and error estimates, but also information about past variations at the location of the transient (up to 30 days in the past) or possible association with a known source from a few external catalogs. Since 2019/11, Fink\footnote{\url{https://fink-broker.org}} receives and processes the ZTF public alert stream.
After reception by Fink, alerts go through a series of treatments (science modules\footnote{\url{https://fink-broker.readthedocs.io/en/latest/science/added_values/}}) that try to characterise the event from the factual information contained in the alert using, for instance, machine and deep learning algorithms, but also resorting to external catalogs to determine if the objects is already known. These science modules are built and provided by the community of users, allowing Fink to build a broad knowledge from Solar System science to galactic and extra-galactic science. As of 2023/01/01, Fink has processed more than 110 million alerts from ZTF, and more than 50 million alerts have already received a classification. All processed alerts are available to the community\footnote{\url{https://fink-portal.org}}.

\subsection{Confirmed Solar System objects}\label{sec:sso_confirmed}

A large majority of the transients seen by ZTF and classified by Fink remains in the same position in the sky over the duration of the survey. It is not the case with SSOs as they quickly move over time in the sky and produce alerts along their trajectories. For each exposure, ZTF performs a cross-match between the alert positions and a daily updated Minor Planet Center (MPC\footnote{\url{https://www.minorplanetcenter.net/iau/mpc.html}}) ephemeris file for all known Solar System bodies within a radius of 30 arcseconds using \texttt{astcheck}\footnote{\url{https://www.projectpluto.com/astcheck.htm}}, and returns the closer object if any. The information about the association is stored in each alert packet. In addition Fink deployed a science module that refines the match by: (a) selecting alerts with a matching radius provided by ZTF below 5 arcseconds, and (b) rejecting alerts that are closer to an object from the Pan-STARRS1 
\citep{2016arXiv161205560C,2020ApJS..251....7F} catalog than to the match from the MPC ephemerides. We note that we currently solely rely on these distance criteria, and we do not take into account other association conditions such as the co-linearity with the expected trajectory to not further delay the processing (see Sect. \ref{sec:limitations}). 

Between {2019-11-01} and {2022-12-29} ({755} observation nights), Fink processed {111,275,131} alerts and {15,828,997} alerts were returned by ZTF with a MPC match ({785,221} unique objects). It represents about {62\%} of all confirmed SSO contained in the MPC database at the time of the analysis, making ZTF one of the largest contributor to asteroid detection to date \footnote{\url{https://sbnmpc.astro.umd.edu/mpecwatch/index.html}}. After applying the filtering described above, Fink kept {15,381,246} alerts ({517,611} unique objects) as matching confirmed Solar System objects\footnote{We also identified 44 comets in Fink's database observed by ZTF which are not included in this analysis.}. The distribution of these alerts on the sky is shown in Fig. \ref{fig:skymap}, and as expected they are mostly located around the ecliptic plane. The median night contains {17,681} alerts associated with confirmed Solar System objects, with a minimum of {29} alerts per night and a maximum of {77,832} alerts per night. These variations are mostly due to the visibility of the ecliptic plane from the ZTF observing site, but also the cadence of the telescope.

\begin{figure}[h]
\includegraphics[width=9cm]{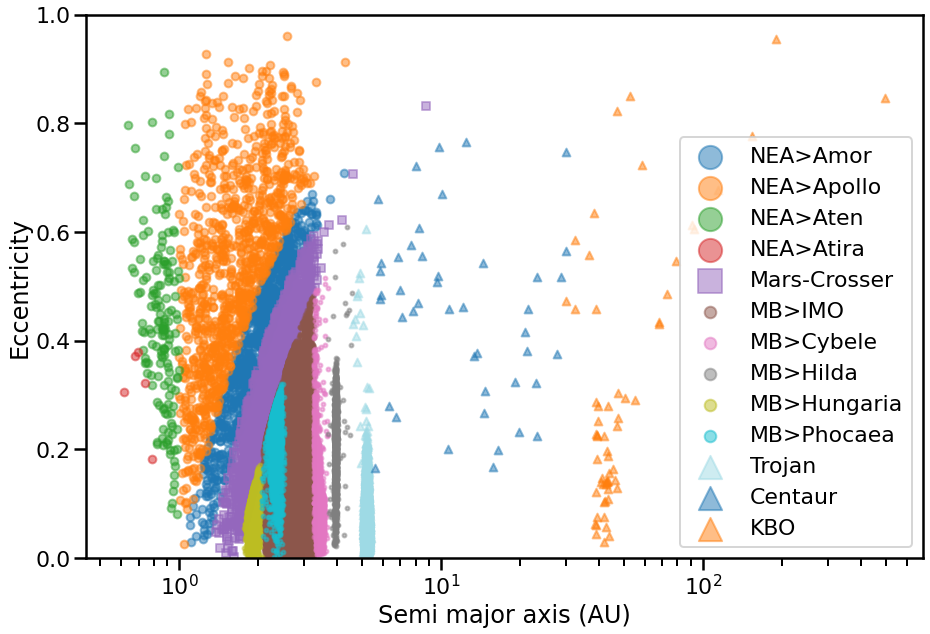}
\caption{Orbital distribution of the 517,611 confirmed Solar System objects in Fink, collected from the ZTF alert stream between 11/2019 and 12/2022. {Objects are color-coded by their dynamical class as defined in the ssoBFT table \citep{2023A&A...671A.151B} as of March 2023}. {Markers denote groups: near-Earth asteroids (NEA, large circle), Mars crosser (square), main-belt (MB, small circle), and outer Solar System objects (triangle). Note:\ MB$>$IMO stands for inner, middle and outer objects from the main belt.}}
\label{fig:sso_confirmed}
\end{figure}

These data set allow us to recover the orbital parameters of the asteroids and, thus, place constraints on the orbit types of the asteroids. \review{For a review of the physical properties of asteroids from ZTF alert data, we refer to \cite{Trilling_2023}.} Overall, ZTF is able to detect a wide range of asteroids from near-Earth (about 1\%) to main-belt (more than 90\%) and trans-Neptunian (a few \%) asteroids. For reference, Fig. \ref{fig:sso_confirmed} displays the distribution of eccentricities of confirmed Solar System objects as a function of their semi-major axes. Each Solar System object generates from one up to more than hundreds alerts over the duration of the survey. This data set is also used to derive constraints on the parameters used in Fink-FAT to later perform the alert association, as reported in Table \ref{tab:sso_condition_inter}. As we show later in this paper, the trajectories are reconstructed assuming a maximum time window between two subsequent measurements (see Sect. \ref{sec:time_window}). We applied this time window when estimating constraints on the parameters of Fink-FAT (\review{defined in Sect. \ref{sec:conditions}).} The parameter values are derived from the 90th percentile on their cumulative distribution and for cadence reasons, we provide different set of parameters for the inter-night and intra-night cases. \review{Intra-night parameters are normalised to one day for all alerts in the night, and the co-linearity condition using $r_\alpha$ is not checked for intra-night trajectories (tracklets).} We note that we are not taking into account the orbit types; hence, this study is mainly driven by the population of main-belt asteroids detected by ZTF which are the most numerous (see also Appendix \ref{app:nea} for further discussion). 
{Furthermore, the parameter values derived from these distributions tend to be more stringent than typical values derived from the literature \citep{2018A&A...609A.113C}, but the rates are not only related to the dynamics of each population; however, they should also be interpreted in the light of instrument capabilities and its cadence, with two subsequent measurements often separated by a couple of days. The 90th percentile threshold was set to minimize the false association numbers while keeping a large number of objects for the analysis.}

\begin{table}
\centering
\caption{Parameters derived from ZTF alerts corresponding to confirmed Solar System objects and used in Fink-FAT to associate alerts between different nights and form trajectories.}
\begin{tabular}{ll}
\hline\hline
Inter-night & \\
$r_d$ &  0.3 deg/day \\
$r_m$ (from same filter bands) & 0.1 mag/day  \\
$r_m$ (from different filter bands) & 0.5 mag/day  \\
$r_\alpha$ & 1.0 deg/day \\
\hline\hline
Intra-night & \\
$r_d$ &  0.03 deg \\
$r_m$ (from same filter bands) & 0.2 mag  \\
$r_m$ (from different filter bands) & 0.8 mag  \\
$r_\alpha$ & -- \\
\hline
\end{tabular}

\label{tab:sso_condition_inter}
\end{table}

\subsection{Solar System object candidates}\label{sec:sso_candidates}

\review{Between {2019-11-01} and {2022-12-29} ({755} observation nights), Fink processed {111,275,131} alerts and {5,807,587} alerts were sent by ZTF with a single measurement or with up to two detections separated by less than 30 minutes, from positive subtraction with the reference image, and without a match with the MPC database. This is what we would naively get in input of a linking code for example without any other treatment.} The Fink science module that returns confirmed Solar System objects also provides information about new Solar System object candidates. 

An alert is considered as such a candidate if it satisfies the following criteria: 1) the alert is not matched to a confirmed Solar System object; 2) the alert is a newly detected object, or it has up to two detections separated by less than 30 minutes; and 3) the alert is not close to a star-like object ({using the star-galaxy separation score,} sgscore1 $< 0.76$) from the Pan-STARRS1 catalog (distance below $ 5''$).

Within the same period of time, {389,530} alerts have received the Solar System candidate tag, with a median of {308}  alerts per day, a minimum at {1}  alerts in a night and a maximum at {12,889} alerts in a night. We note that the distribution varies over time, but broadly follows the distribution of confirmed Solar System objects. The location on the sky of the alerts satisfying the previous criteria is shown in Fig. \ref{fig:skymap}. We can see a excess {along the ecliptic plane at zero right ascension and declination (albeit two orders of magnitude smaller than the confirmed objects)}, but there are also dense regions further away.

The SSO module gives a first estimation of the nature of an alert. However, this first guess can quickly turn up to be wrong as new incoming alerts are processed. Of the {389,530} alerts initially associated with Solar System candidates, {3,772} have been associated with another alerts at the same location on the sky emitted the next nights ($\sim$1\%). These erroneously classified objects were mostly found later to be extra-galactic (e.g., supernova candidates) or remained unclassified. \review{All Solar System candidate alerts can be accessed using the Fink REST API, see App. \ref{sec:api}.}

\section{Validation of confirmed Solar System objects} \label{section_results}

\review{Each night, Fink extracts about 300 new Solar System candidates (median), and 18,000 confirmed Solar System objects (median). Since Fink-FAT will be applied on Solar System candidates only during operations (not the confirmed ones), if we run Fink-FAT on ZTF confirmed Solar System objects, this would basically mean a factor 60 in data volume; this is in line with what we expect with LSST in terms of data volume (or, rather, pessimistic). Therefore in this section, we use the confirmed Solar System objects data set to test the performances of Fink-FAT, both in terms of technical capabilities and scientific results.}

For this test, we used a subset of all the ZTF alerts associated with confirmed Solar System objects running from 2020-09-01 to 2020-10-01 (24 observation nights). This period was chosen based on the large number of confirmed Solar System alerts: {796,486} alerts in total with a median of {26,993} alerts per night, a minimum of 3,314 alerts, and a maximum of 69,831 alerts. This high volume of alerts per night allows us to also test Fink-FAT with a number of alerts close to the expected LSST flow rate for the Solar System object candidates, which is essential as one of our objectives is to overcome the data rate challenge of the LSST\footnote{\url{https://lse-163.lsst.io}}.

In the following, all tests were performed on the Fink Apache Spark Cluster deployed on the VirtualData cloud. The cluster makes use of Intel Core processors (Haswell architecture) at 2.3 GHz. The association algorithm is fully sequential, so it uses only one core during its execution, but it has access up to 36 GB of RAM. The orbit\ fitting however is deployed on a cluster of machines with the following configuration: a total of 24 cores split in four cores per executor (so six executors) and 8 GB of RAM per executor.

\subsection{Time performance}\label{sec:time-perf}

The first experiment with Fink-FAT was to determine the computation time for the association and orbit fitting steps. On average, Fink-FAT took {77} seconds (median) to perform the association step each night. The minimum association time was 8 seconds and the maximum was {261} seconds. The median trajectory volume sent to \orbfit\ each night was {3,543}, the minimum was 7 and the maximum was {10,334}. The orbit fitting step took on average {291} seconds each night (median), with a minimum execution time of {35} seconds ({7} trajectories), and the maximum of {744} seconds ({10,334} trajectories). The total execution time for the entire month of data (24 nights) on 24 cores was about {168} min. {The orbit fitting step takes a significant part of the total computation time with about 119 minutes (70.83\%), while the association step takes about 40 minutes (23.81\%) and the time taken to retrieve all the alerts from Fink database is about 10 minutes (5.95 \%).}

\begin{figure}[h]
    \centering
    \includegraphics[width=0.5\textwidth]{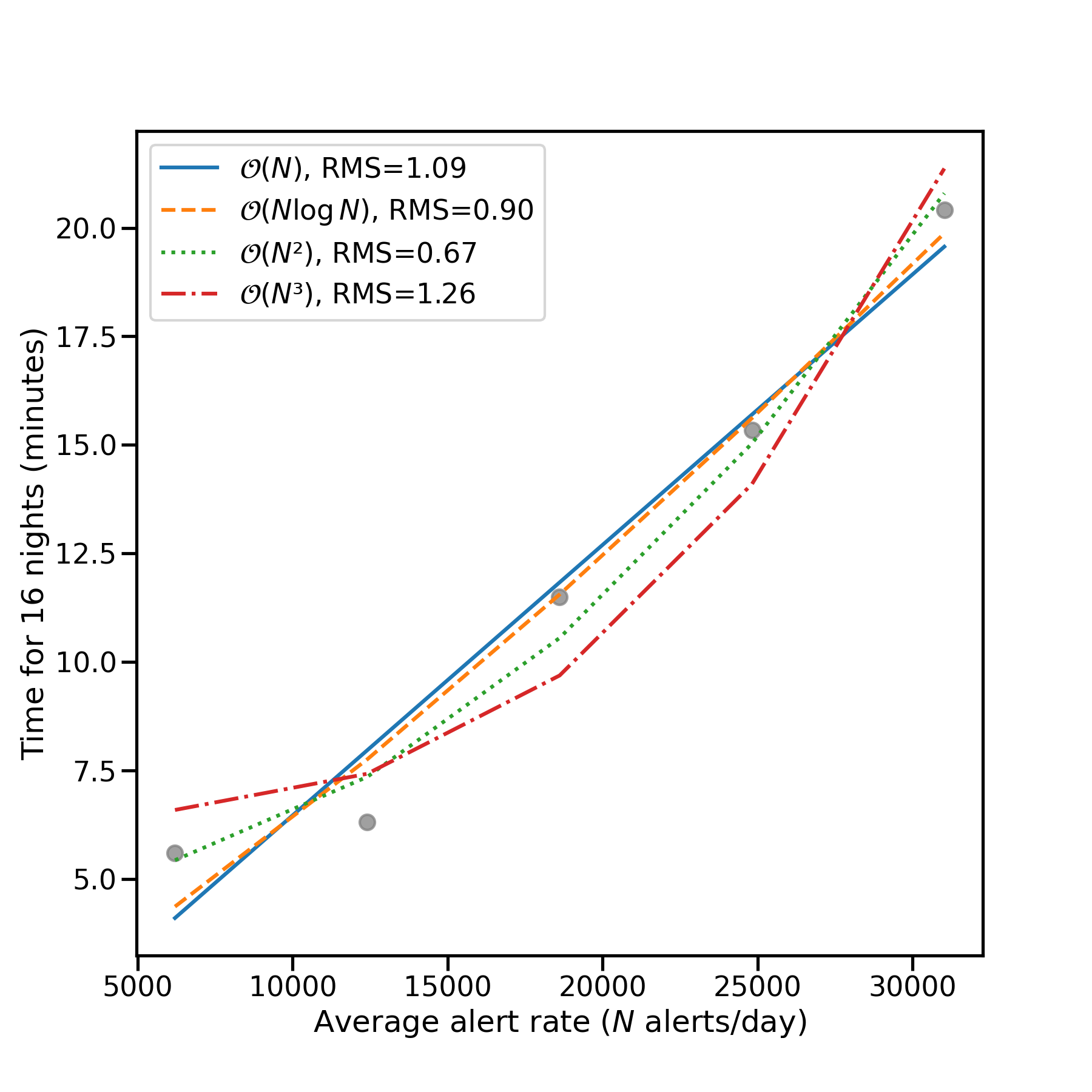}
    \caption{Computational time taken by the association step as a function of the nightly Solar System  alert rate (average), for 16 consecutive nights of ZTF alert data. Various functions have been fitted to the data to give a hint on the run-time complexity of the algorithm described in Sect. \ref{sec:assoc_steps}, with the root mean square value displayed in the legend.}
    \label{fig:fink_fat_comp}
\end{figure}

\review{In order to explore the complexity of Fink-FAT, we ran several experiments. First we decoupled the association step (described in Sect. \ref{sec:assoc_steps}) and the orbit fitting step (depending on the \orbfit\ software; see Sect. \ref{sec:orbfit_description}).}

\review{For the association step, we chose a period of 16 consecutive observing nights\ and we varied the number of Solar System alerts sent each night to Fink-FAT from 6,000 alerts/night to 30,000 alerts/night by sampling the number of Solar System alerts each night. Figure \ref{fig:fink_fat_comp} shows the computational time as a function of the alert rate (grey circles), with fixed allocated resources (single core, with up to 35GB RAM). As the alert rate increases, the time increases. We approximate the run-time complexity of the algorithm by fitting multiple functions (linear, linearithmic, quadratic, cubic) to the data. The best-fitted function (the smallest root mean square value) has a quadratic dependency in the number of alerts per night, which makes Fink-FAT no better than current algorithms (for example HelioLinC \citep{2018AJ....156..135H} is linearithmic). Such a complexity is probably not very encouraging as such; however, regarding the computation time reaching a maximum of about 20 minutes over a time window of 16 nights with on average 30,000 alerts per night (expected Solar System alert rate for LSST), we conclude that it is already fast enough to be used in the context of the forthcoming LSST survey.\footnote{We note though that the total time can vary within a factor of 1-1.25, depending on the load of the cloud platform used. However,  this always remains fast enough overall.}}

\review{For the orbit fitting step, where the computation is  straightforwardly parallel and distributed over many machines, we performed two experiments. First, we fixed the allocated hardware resources (16 cores, 2 cores per executor, and 2 GB of RAM per core) and we recorded the computation time as a function of the number of trajectories generated by the association step. We varied the number of trajectories from 100 to 10,000 and we observed a linear increase of computational time. Second, we set the number of trajectories to 5,000 and we recorded the computation time as a function of the number of allocated cores (from 2 cores to 128 cores). As expected, the computational time is inversely proportional to the number of cores allocated in the range of resources allowed. This behaviour is encouraging as, even if the orbit fitting step is taking most of the computation time of Fink-FAT, it scales linearly with the allocated resources.}

\subsection{Reconstruction performance} \label{sec:finkfat-recov-perf}

\begin{table}
\centering
\caption{Performance of Fink-FAT on the reconstruction of the confirmed Solar System objects between 2020-09-01 and 2020-10-01.}
\begin{tabular}{lcc}
 & \multicolumn{2}{c}{Fink-FAT} \\
\hline\hline
a. Confirmed objects & \multicolumn{2}{c}{{87,076}}  \\
b. Detectable objects & \multicolumn{2}{c}{{43,919}}  \\
\hline
& All orbits & Orbit with errors \\
c. Reconstructed Orbits & {39,628} & {13,252}   \\
d. + Pure  & {28,719} &  {12,853}  \\
e. + Unique & {19,956} & {10,755}  \\ 
\hline
Purity (d/c)              & {72.5} \% & {97.0} \%  \\
Efficiency (e/b)         & {45.4} \% &  {24.5} \%  \\
\hline
\end{tabular}
\label{tab:finkfatperf}
\end{table}

In this section, we explore the performance of Fink-FAT in correctly reconstructing trajectories. The results are summarised in the Table ~\ref{tab:finkfatperf}. \review{The first two lines are the description of the input dataset: \textbf{a.} the number of confirmed Solar System objects; and \textbf{b.} the number of detectable objects (Sect. \ref{sec:detectable}); \textbf{c.} gives the number of reconstructed orbits, that is, the set of trajectories for which the orbit fitting step returns valid orbital elements (Sect. \ref{sec:detectable}); \textbf{d.} and \textbf{e.} show the number of pure reconstructed orbits and unique reconstructed orbits, respectively (Sect. \ref{sec:pure_and_unique}). Finally, we show the purity and the efficiency as two metrics to assess the efficiency of the method. Each line also contains the number of corresponding orbits with valid error estimates, {that is with full differential corrections applied}.}

\subsubsection{Detectable and reconstructed orbits} \label{sec:detectable}

There are {87,076} confirmed Solar System objects in the test dataset, and {43,919 (50.44\%)} are detectable by Fink-FAT. We defined two conditions to establish a detectable trajectory by Fink-FAT: (1) the trajectory must have a number of alerts greater or equal to the minimum number of alerts required to be processed by \orbfit and (2) the number of separating nights between each alert must be less than the time window parameters (see Sec \ref{sec:time_window}). For this test, the minimum number of alerts for \orbfit\ was six, and the time window was set to fifteen days. 

After the association and the orbit fitting steps, Fink-FAT output {39,628} trajectories with valid orbital parameters from the detectable trajectories (i.e., initial orbit determination was successful). The longer trajectories was made of {12} alerts, and approximately 50\% of the trajectories had the minimum of six alerts. A large part of the trajectories ($\sim${80.3\%}) starts with an intra-night association or a pair of alerts from different nights ($\sim${12.3\%}). The remaining trajectories begin with the association of an old alert with an intra-night association (see Fig. \ref{fig:association_diagram}).

\subsubsection{Pure and unique orbits} \label{sec:pure_and_unique}

{Each step of the association algorithm can produce internally different trajectories including the same alert. Hence,} some trajectories in the sky may spuriously intersect when fitting for orbits. Therefore, we defined the pure orbits as the trajectories containing only the observations of the same Solar System object. Fink-FAT returned {28,719} pure orbits. We define the purity of Fink-FAT outputs as the ratio between the number of reconstructed orbits and the pure orbit, which is about {72.5 \%} for this dataset. In addition, multiple disconnected trajectories can come from the same Solar System object. It is a direct consequence of the time window and the \orbfit\ limit parameters. By taking only unique Solar System identifiers, Fink-FAT returned {19,956} asteroids. We define the efficiency of Fink-FAT as the ratio between the number of detectable SSO and the uniquely detected SSO, which is {45.4 \%} for this experiment.

{Finally as the observational arcs are small, the orbit fitting procedure does not always fully converges. In the case where only the initial orbit determination is available, we have a set of orbital parameters without associated errors (hence, it is rarely accurate, but often enough for short term predictions), while if the full differential corrections step has succeeded we have a better estimation on the orbital parameters that includes the estimated covariance for the parameters (hereafter, orbits with errors).} From Table \ref{tab:finkfatperf}, Fink-FAT reconstructs 39,628 orbits, but only 13,252 pass the full differential correction step and have errors in their parameters (33.44 \%). However, the ratio between the number of reconstructed orbits with an error and pure orbits (purity) with an error is almost 97 \%. This means that despite the relatively low efficiency, if we have an orbit with an associated error estimate, we are almost certain that this orbit is valid, which is a crucial information when planning follow-up observations.

\begin{table}
\centering
\caption{Detection performance of Fink-FAT \review{between 2020-09-01 and 2020-10-01} by orbit dynamical classes.}
\begin{tabular}{lrr}
 & {Initial orbit distribution} & \review{Reconstructed}  \\
\hline\hline
Cybele & {172} & {56} \textcolor{gray}{({32.56\%})}  \\
Main belt & {40,533} & {10,229}  \textcolor{gray}{({25.24\%})}  \\
Phocaea & {401} & {97}  \textcolor{gray}{({24.19\%})} \\
Jupyter trojan & {1041} & {198} \textcolor{gray}{({19.02\%})}  \\
Hilda & {186} & {33}  \textcolor{gray}{({17.74\%})}  \\
Mars crosser & {455} & {46} \textcolor{gray}{({10.11\%})} \\
Hungaria & {700} & {63}  \textcolor{gray}{({9.00\%})}  \\
Amor & {51} & {2}  \textcolor{gray}{({3.92\%})}  \\
Apollo & {48} & 1  \textcolor{gray}{{(2.08\%)}}  \\ 
KBO & {8} & {0}  \textcolor{gray}{({0.00\%})}  \\
Aten & {5} & {0}  \textcolor{gray}{({0.00\%})} \\
Centaur & {3} & {0}  \textcolor{gray}{({0.00\%})} \\
Atira & {0} & {0}  \textcolor{gray}{({0.00\%})} \\
\hline
\end{tabular}
\label{tab:perforbittype}
\end{table}

\subsubsection{Orbit types}

\review{Table \ref{tab:perforbittype} shows the detection performance of Fink-FAT by orbit dynamical class. The first column displays {orbit dynamical classes from the ssoBFT table \citep{2023A&A...671A.151B} as of March 2023} and present in the test dataset. The second column shows the number of detectable Solar System objects per orbit class in the test dataset. The third column displays the number of pure and unique reconstructed orbits with error estimates recovered by Fink-FAT. The percentage recovery with respect to the initial orbit distribution is shown in parenthesis in grey.} The best-reconstructed objects are, not surprisingly, the objects from the main belt (MB, Hungaria, Phocaea, Hilda) and the Jupiter trojan as the Fink-FAT association parameters were derived mostly from main-belt objects. On the other hand, the closest and the farthest objects are not detected. \review{The almost zero efficiency for NEO and KBO is directly related to the reason behind the overall low efficiency. Near-Earth asteroids (Amor, Apollo, Aten, Atira) and KBO associations would have occurred in later steps in the association pipeline (mainly in the last step, when we associate single measurements from different nights together), but their elements were already discarded by previous association steps.}

We note that the sum of the "initial orbit distribution" column in Table \ref{tab:perforbittype} does not match the number of detectable objects in Table \ref{tab:finkfatperf} due to a mismatch in names between ZTF and MPC. The difference between the two is 316 objects. The asteroids can have up to four identifiers in the MPC database (number, name, principal designation, and other designations) that we use for the correlation, but as the MPC database is frequently updated, names can change over time. {To reduce the confusion, Fink-FAT is now using the Virtual Observatory Solar System Open Database Network (SsODNet) services \citep{2023A&A...671A.151B}, notably available from \texttt{rocks}\footnote{\url{https://rocks.readthedocs.io}}}

\begin{figure}[h]
\includegraphics[width=9cm]{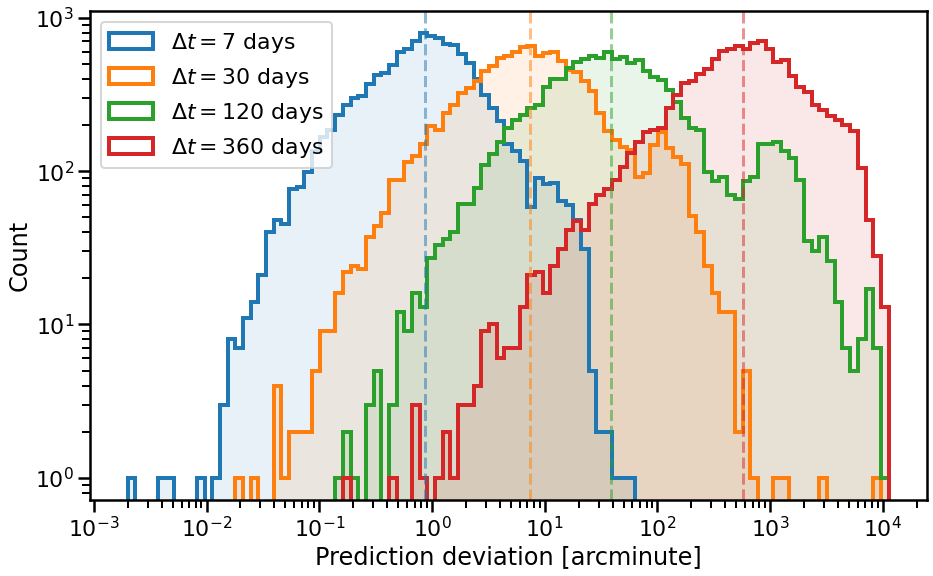}
\caption{Histogram for the deviation (in arcminute) between the ephemerides generated using the orbital parameters estimated from Fink-FAT trajectories (pure trajectories from full orbit determination), and the ephemerides generated using the orbital parameters taken from MPC for the corresponding objects. We vary the time from the last observation to the computed ephemeris for each trajectory: $\Delta t = 7$ days after the last observation (blue), $\Delta t = 30$ days (orange), $\Delta t = 120$ days (green), and $\Delta t = 360$ days (red). The median of each distribution is shown as dashed vertical line.}
\label{fig:deviation_follow_up}
\end{figure}

We also used the cross-match with the MPC orbit database to assess the quality of the orbits computed by \orbfit. For each orbital parameter, the median of the residue distribution was below 1$\%$. The best reconstructed orbital parameters are the semi-major axis, eccentricity and inclination. As expected, the three others parameters (longitude of the ascending node, argument of periapsis, and mean anomaly) had a long tail in their residue distribution, due to the small number of observations per object input to \orbfit\ (and the corresponding arcs have a median of nine days). In order to translate this residue in terms of useful information for the follow-up of these objects, we computed the deviation (in arcminute) between the ephemerides generated using the orbital parameters from Fink-FAT pure and unique trajectories, and the ephemerides generated using the orbital parameters from MPC for the corresponding objects, after several days from the last observation of each trajectory. The results are displayed in Fig. \ref{fig:deviation_follow_up}. Seven days after the last observation of each trajectory, the median deviation between the predictions is about 1 arcminute. This means for any follow-up telescope with a field of view greater than 1 arcminute, most of the objects should be detectable by pointing to Fink-FAT predictions. However, as time goes on (and assuming no new observations are added to Fink-FAT), the median deviation between Fink-FAT predictions and the predictions from the MPC-based orbital parameters increases: 7 arcminutes after 30 days, 38 arcminutes after 120 days, and 577 arcminutes (9.6 degrees) after one year. 
This means that without any new information, Fink-FAT predictions on object trajectories can be considered as useful for follow-up observations over a month \review{(note: the initial arc lengths used for predictions have a median value of nine days).}

\subsection{Time window impact}\label{sec:time_impact}

In the previous sections, we set fixed the time window parameters to associate alerts when forming trajectories: the separation time between the end of a trajectory and a new alert was set to 15 days, the time to keep an old alert as candidate was set to 2 days, the time to keep an intra-night as candidate was set to 2 days. We also increased these time window parameters to assess the impact on the orbit recovery. We observe a decrease in efficiency when the time windows increase. During the experiments with the largest time windows, the association step generated a larger number of trajectories than the baseline case, but fewer trajectories ended with orbital elements in the orbit fitting step. The reduction in efficiency was explained by a higher rate of false positives (especially in the pure orbit step) as many trajectories were crossing each other due to high density of objects from the main belt near the ecliptic plane.

\subsection{Comparison with present linking software} \label{sec:comparison}

\review{To consider a trajectory as detectable, Fink-FAT requires a minimum of six observations with no more than 15 days between two observations. Compared to MOPS or HelioLinC which rely on tracklets, this strategy gives more flexibility with respect to the choice of cadence for a survey. However, in practice, Fink-FAT performances still rely heavily on the presence of tracklets, which makes it more prone to cadence effects than purely tracklet-less algorithm such as THOR.}

\review{The efficiency of Fink-FAT, defined as the ratio between the number of detectable SSO and the unique detected SSO remains rather low (25-45$\%$). We note that this result is obtained on ZTF observations, including all the real life effects such as unequally spaced cadence. THOR \citep{moeyens2021thor} on a similar dataset (ZTF alerts from 2018), but with a different criterion to define detectable trajectories (five observations instead of six in the case of Fink-FAT), reports a overall completeness for the main-belt objects and beyond of 97.4 \%, while other works (e.g. \cite{2018AJ....156..135H}) also report high efficiencies despite different detectable definition. The low efficiency for Fink-FAT can mainly be explained by the fact that the alert association steps are sequential: previous steps will remove the associated elements (trajectories, intra-night trajectories, or single alert) from the possible association for the next steps (see Sect. \ref{section_fink_fat}). Hence, a true association that would show up only in a later step could never be considered because its elements would have been mismatched to other elements in a previous step.}

\review{The purity reached by Fink-FAT is as high as 97 \% after full orbit fitting. This is comparable to what THOR and others currently report. This result is encouraging as, while Fink-FAT is missing many of detectable objects, it provides a low rate of false trajectories, which is crucial when optimizing the limited follow-up time, for example.}

\review{For the set of parameters chosen, Fink-FAT computational performances are dominated by the orbit fitting step (see Sect. \ref{sec:time-perf}), and not the association steps. This is mainly due to the fact that the association steps are applied sequentially (with the same fact giving rise to the low efficiency). The end-to-end running time (for equivalent computing resources) from associating alerts to extracting orbital parameters is lower than other (more precise) software. For example, in the previous experiment using one month of ZTF alert data, Fink-FAT returned full results in about {168} minutes on six nodes of four cores each, while THOR, based on two weeks of ZTF alert data, reported a computational time of about 18 hours using 23 nodes with 28-cores per node. This represents a factor of $\sim$ 350 in speed-up (assuming linear scaling with the data volume for THOR). We note though that for an extended choice of parameters (i.e., giving more flexibility to associate elements), we observe a degradation of the Fink-FAT computational performances by a factor of 5 (see Appendix \ref{app:nea}).}

\section{Application on candidate Solar System objects} \label{section_discovery}

In this section, we describe how we applied Fink-FAT on the set of Solar System object candidates from Fink. We also report the results from two follow-up campaigns performed to further validate the results.

\subsection{Reconstructed orbits}

Fink database contains 389,496 alerts classified as Solar System candidates between 2019-11-01 and 2022-12-29. These alerts were not matched with the minor planet ephemerides generated from MPC at the time of the observations and we provide them to Fink-FAT for the association and orbit fitting. While the total number of observations is comparable to the number of confirmed objects used to validate Fink-FAT (one month of data, see Sect. \ref{section_results}), the nightly rate becomes much smaller as the time spanned is greater, with a median rate of 292 alerts per night, a minimum of 0 alert (only one night) and a maximum of 12,889 alerts per night. 

We give to Fink-FAT the same parameters as the previous experiences done with the confirmed Solar System objects. Fink-FAT took 138 minutes to finish its computation over the three years of Fink's data. The time to associate the alerts became the shortest (9 minutes) compared to the other tests, and the request time is no longer negligible (39 minutes). The orbit fitting is still the most significant part of the computation time (90 minutes). This experiment uses the same hardware configuration than the experiments with the confirmed asteroids, except the orbit fitting, which is performed locally on three cores as the volume of data is small. 

Fink-FAT sucessfully linked 2,025 observations (0.5\% of all the candidates) to form a total of 327 trajectories with an orbit estimate, including 182 orbits with error estimates on the orbital parameters (55\%). Overall, 271 trajectories have six measurements (83\%) and the longer trajectory (only one) has nine measurements. The distribution on the sky of these alerts is shown in Fig. \ref{fig:skymap}, and they are all located around the ecliptic plane, at zero declination. 

\begin{figure*}[h]
\includegraphics[width=\textwidth]{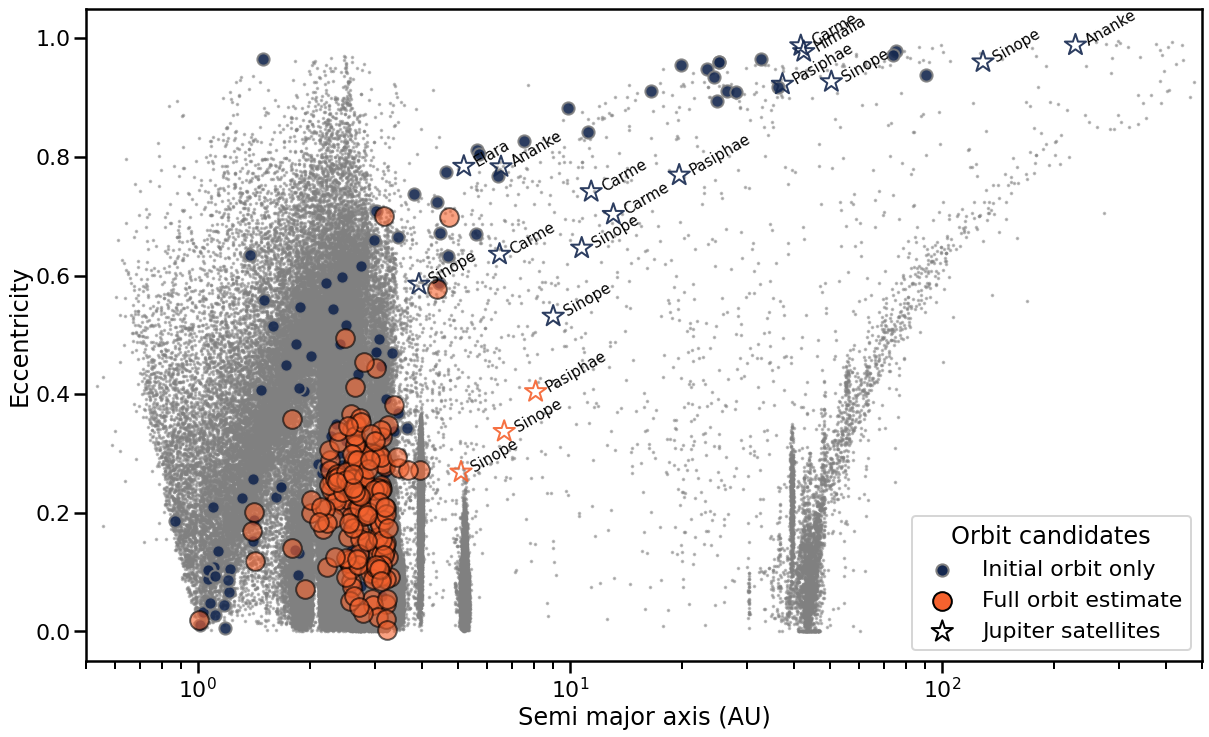}
\caption{Distribution of the 327 orbit candidates returned by Fink-FAT. The orbit candidates that only pass the initial orbit determination step for orbit fitting are shown with dark blue circles. The orbit candidates that also successfully pass the full orbit determination are shown in orange circles. In addition, we show orbit candidates that were later associated with Jupiter satellites with star symbols (see Sect. \ref{sec:jupiter_moons}). For reference, we overplot in grey all the objects from the MPC database as of March 2023.}
\label{fig:sso_futur}
\end{figure*}

The distribution of magnitudes of the alerts in the trajectories linked by Fink-FAT is similar to the distribution of magnitudes for confirmed Solar System objects. The distributions of the orbital parameters and error estimates follow the same trend as for the confirmed and pure orbits described in Sect. \ref{sec:finkfat-recov-perf}. Hence, according to Table \ref{tab:finkfatperf}, this points towards a high purity and it lends confidence to the fact that the orbit candidates with error estimates might be valid unreported Solar System candidates in the MPC database at the time of the observations. In Fig. \ref{fig:sso_futur}, we show the distribution of orbital parameters estimated from reconstructed trajectories. \review{Trajectories that pass the full orbit determination are mainly located in the main belt, while those from only initial orbit determination tend to lie more often on extreme regions of the parameter space, {with a perihelion at 1 AU, which is likely a sign of ill-defined orbit solutions driven by the initial conditions used in the solver.} This is probably a consequence of the fact that Fink-FAT linkage parameters estimated from the set of confirmed objects are mainly representative of main-belt objects (see Sect. \ref{sec:sso_confirmed}).}

\subsection{Accounting for updates}

When selecting the Solar System object candidates, we rely on the fact that ZTF did not find any counterparts when crossmatching with the ephemerides provided by the MPC. In addition, we did not attempt to check for data elsewhere when associating alerts to form trajectories. Yet as more observations are performed, more Solar System objects are discovered and eventually added to the MPC database or available somewhere else. Therefore, to check if any of our alerts from candidate trajectories could be associated with a currently known asteroids, we perform an association by ephemerides with the SkyBot cone-search tool \citep{2006ASPC..351..367B} with an up-to-date version of the Solar System object data. To perform the association, we used a cross-match radius up to five arcseconds between the SkyBot predictions and candidate alerts, {as well as a threshold on the variation with respect to the predicted magnitude at 0.3 mag}. 

\begin{figure*}
    \centering
    \includegraphics[width=0.3\textwidth]{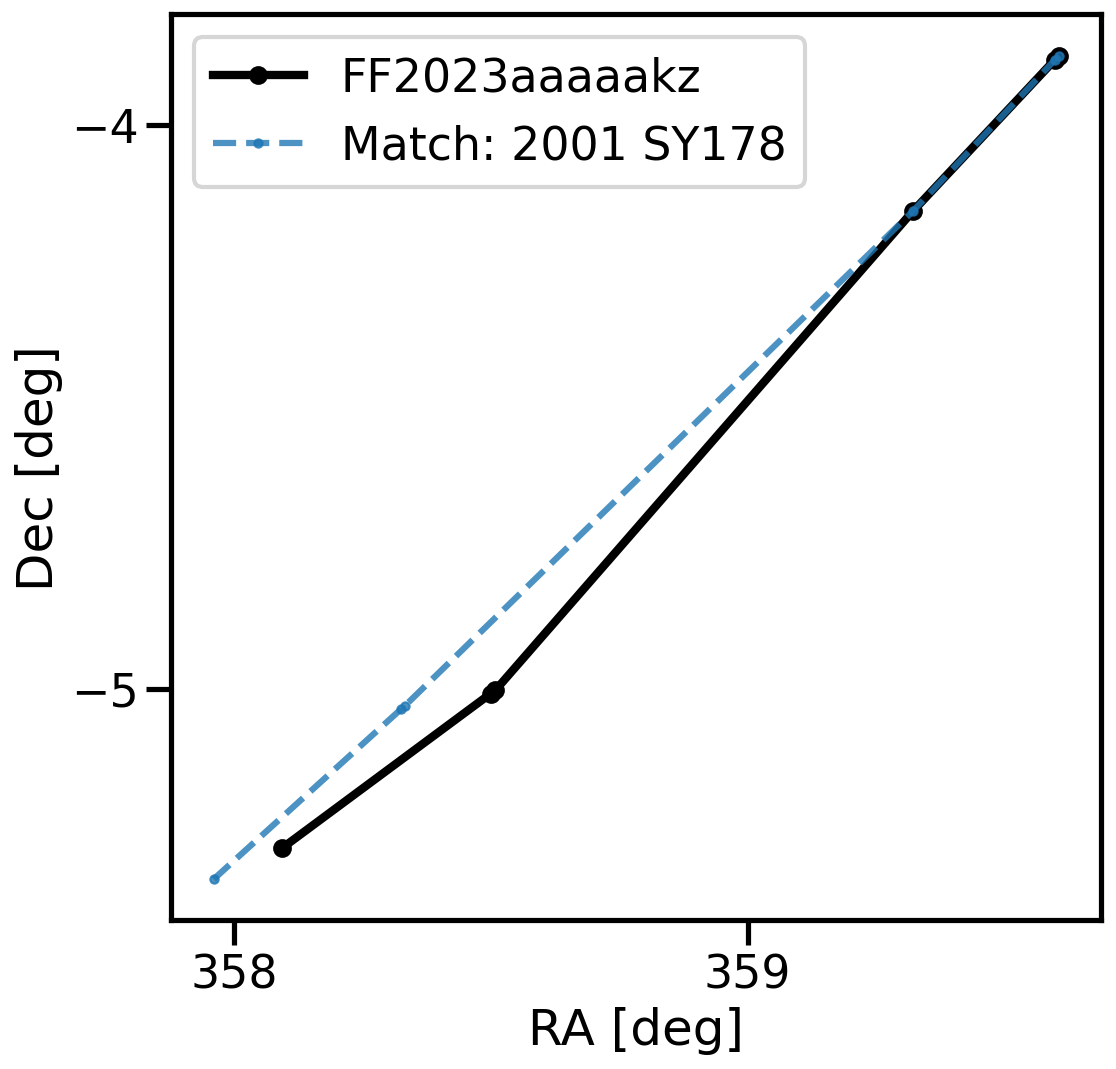}
    \includegraphics[width=0.3\textwidth]{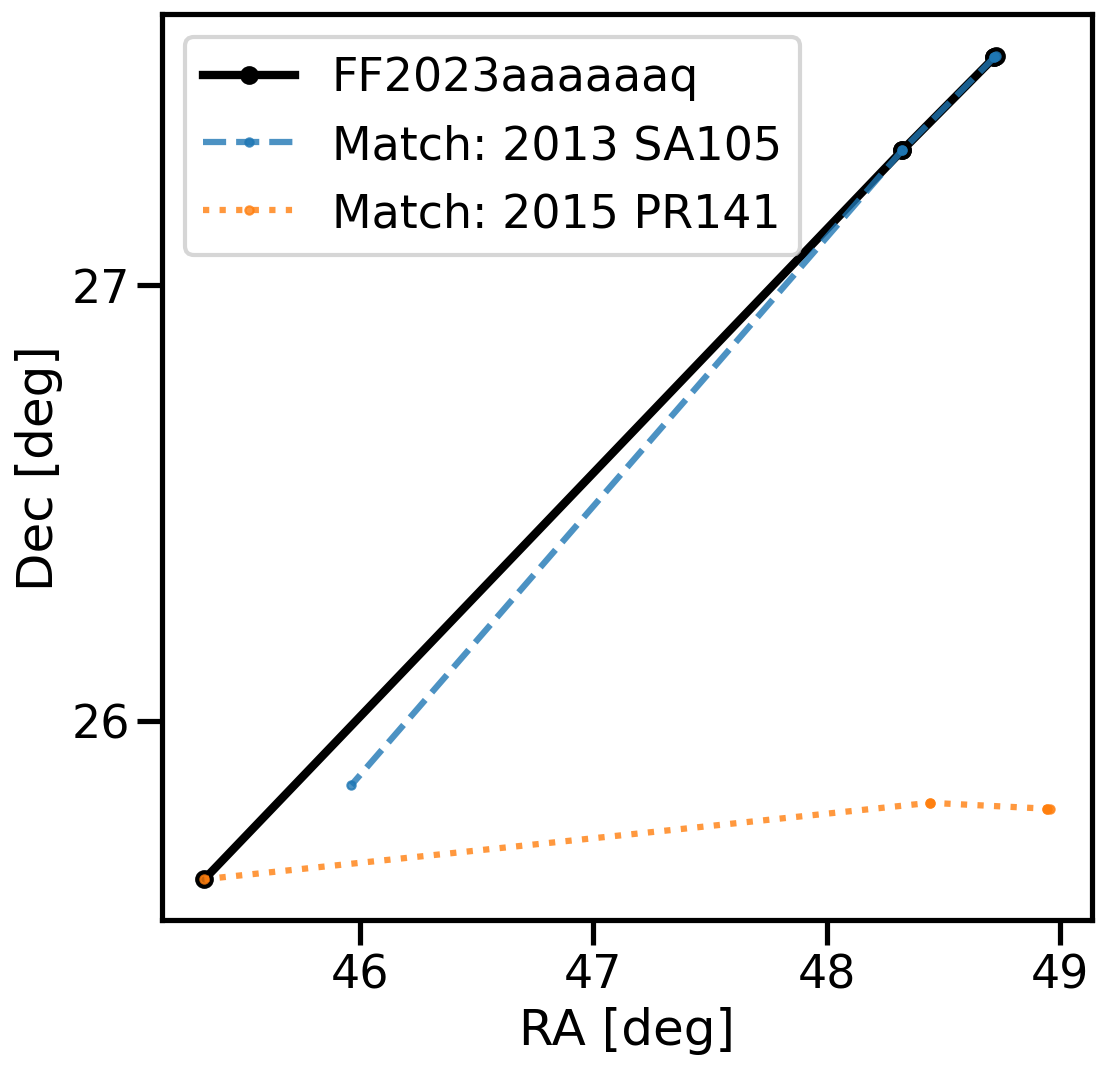}
    \includegraphics[width=0.3\textwidth]{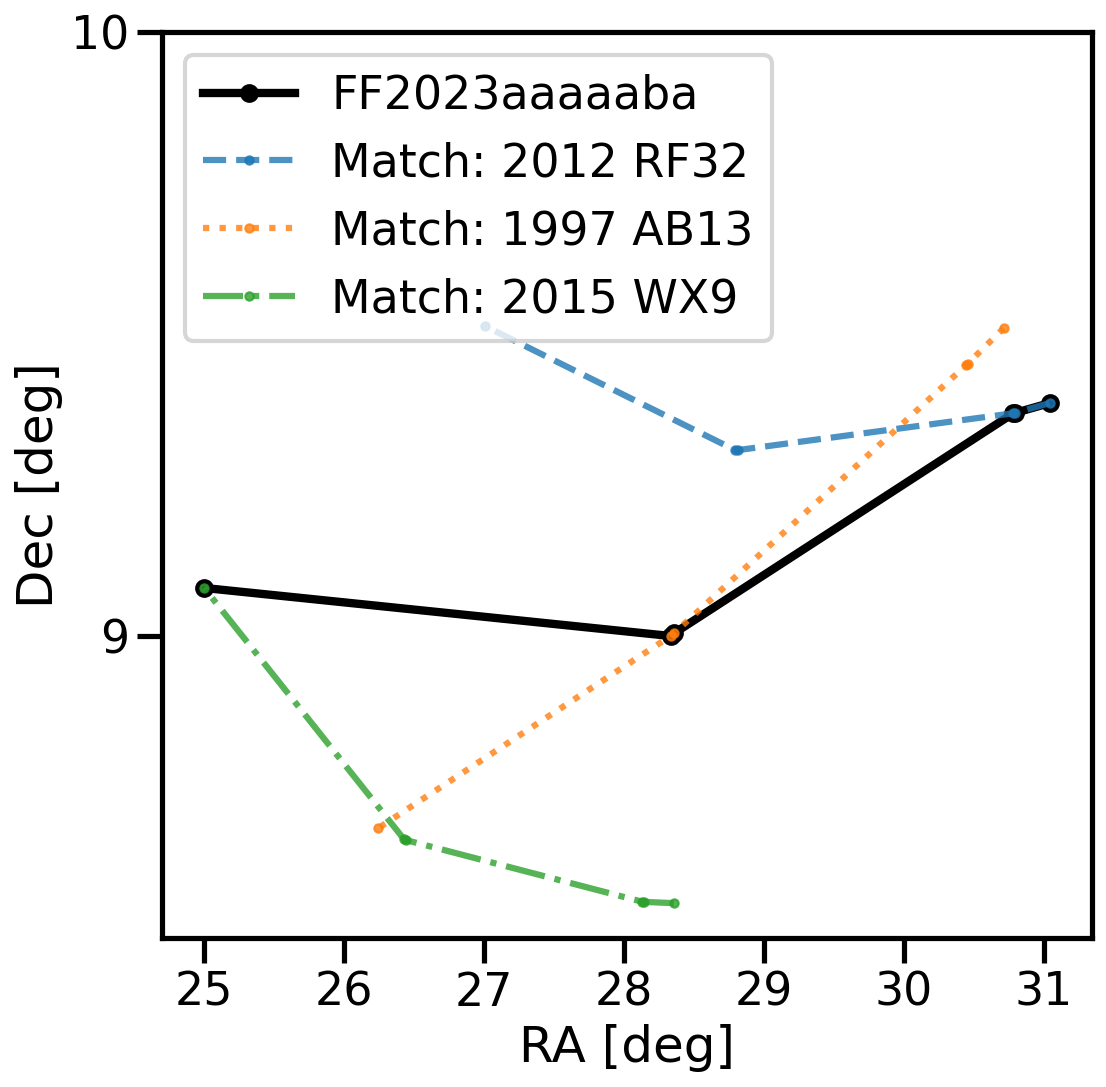}
    \caption{Examples of spurious trajectories returned by Fink-FAT in the RA-Dec space. In all panels, the initial trajectory is in solid black line. \texttt{FF2023aaaaakz}: the two top-right corner alerts were matched to 2001 SY178, but the epheremides of this object is not compatible with the position of the remaining alerts (which are at about 5.2 arcseconds from 2004 NE13). \texttt{FF2023aaaaaaq}: the top-right corner alerts were matched to 2013 SA105, while the bottom left alert was matched to 2015 PR141. \texttt{FF2023aaaaaba}: the top-right corner alerts were matched to 2012 RF32, the middle alerts (intra-night) was due to the passing of 1997 AB13, and the middle left alert was matched to 2015 WX9. 
    }
    \label{fig:candidates_trajectory_examples}
\end{figure*}

We found {1,284 (63\%)} alerts with a previously unreported counterpart. Out of the 327 candidate trajectories that pass the orbit fitting, {92 (28\%)} had all their alerts associated with the same Solar System object (pure orbit like). {Then, 170 trajectories (52\%)} had associations coming from multiple asteroids (orbit is not pure). In this case, there are two types: trajectories for which most of the observations are matched to the same asteroid (or to no asteroids) but one and the trajectories for which most of the observations are from different asteroids (see Fig. \ref{fig:candidates_trajectory_examples}). Unfortunately, the high density of asteroids in the main belt contributes to this false associations. Finally {65 trajectories (20\%)} were not associated with any known objects and were used for the follow-up campaigns. \review{We note that for most of those composite trajectories, \orbfit\ failed to return orbital parameter error estimates, which is only the initial orbit determination step was successful and we can easily discard them.}

\subsection{Follow-up campaigns}

In order to further validate the candidate trajectories from Fink-FAT, we organised two follow-up observation campaigns using the telescope network of the Las Cumbres Observatory (LCOGT, 1 meter) \citep{Brown_2013} and the Observatoire de Haute Provence (OHP, 1.2 meter), France. The first campaign took place in July 2022 with trajectories candidates detected by Fink-FAT in 2021. The second campaign took place in late September 2022 with candidates trajectories from August 2022. To guide our decision for the follow-up, the trajectories candidates are sorted based on the best error estimate on the three first orbital parameters (semi-major axis, eccentricity, and inclination); however, due to technical problems with the LCOGT northern telescopes at the time of observations, we were restricted to ZTF-derived trajectories visible from the southern hemisphere only which left only few candidates (and these are not necessarily the best). 

\subsubsection{First observation campaign}

Initially, no trajectories were visible from the Cerro-Tololo (W87) site for the first observation campaign (2022-07-05). We decided to increase the time window parameter of Fink-FAT from two days to eight days for inter-night association in order to get candidates and not lost the observing time. Two trajectories were finally visible from the site and one was selected for a follow-up study. The trajectory was detected by Fink-FAT in 2021 (last alert emission date after extension by ephemerides in 2021-05-22, that is, more than a year before the follow-up observations), with an arc of 46 days. The orbital parameters were estimated to (a[AU], e, i[deg]) = (3.0593, 0.22603, 16.66617). The observations confirmed the position of a moving object in the exposure (about 9 arcminutes away from the predicted ephemeris). However this object was already known and contained in the MPC database (MPC number: 525570) with orbital parameters (a[AU], e, i[deg]) = (3.0652517, 0.2243976, 16.77083). The asteroid was unknown in Fink initially because the alerts must fall within 5 arcseconds of a known asteroids to be associated (see Sect. \ref{sec:sso_confirmed}) and it was just beyond the threshold for association ($\sim$ 6 arcseconds). Despite this, it remains a confirmation of the ability of Fink-FAT to detect valid trajectories, but we were rather lucky that the predictions were only 9 arcminutes away from the correct orbit more than a year after the last observations, as according to Fig. \ref{fig:deviation_follow_up}, this object would be in the leftmost tail of the $\Delta t = 360$ days distribution.

\subsubsection{Second observation campaign}

For the second observation campaign, we ran Fink-FAT with its default parameters. Unlike the first campaign, the trajectories were predicted about one month before the follow-up observations, so we would expect deviations in the predictions around a dozen of arcminutes (see Fig. \ref{fig:deviation_follow_up}). We selected six trajectories of six observations each from ZTF observations taken in August 2022. The follow-up data was acquired from the LCOGT site on 2022-09-25 and 2022-10-01 and from the OHP site on 2022-09-26. Five trajectories have received follow-up, three trajectories were found to be Jupiter irregular satellites (J9 Sinope and J8 Pasiphae) and for two, no counterparts were found. In the following, we detail each trajectory and the follow-up observations.

\paragraph{\texttt{FF2023aaaaama}:} the last alert emission date was on 2022/08/28, and the observations were performed on 2022/10/01 from the LCOGT site. The total arc is 6 days, and the orbital parameters were estimated to (a[AU], e, i[deg]) = (8.085766, 0.404250, 4.198385). There were three moving objects nearby the ephemerides predicted from Fink-FAT estimates. Two were known asteroids (2012 XF166 and 549752), whose positions were not compatible with the initial Fink-FAT trajectory. The remaining object was an irregular moon of Jupiter, Jupiter VIII Pasiphae ($\approx$ 23 arcseconds from the Fink-FAT predictions). We found Pasiphae was also compatible with the initial Fink-FAT trajectory ($\leq$ 1 arcseconds distance from all alerts) and we concluded that \texttt{FF2023aaaaama} was an observation of Pasiphae.

\paragraph{\texttt{FF2023aaaaamb}:} the last alert emission date for this trajectory was on 2022/08/28, and the observations were performed on 2022/09/25 from the LCOGT site. The total arc is four days and the orbital parameters were estimated to (a[AU], e, i[deg]) = (6.657587, 0.337133, 2.500486). There were three moving objects nearby the ephemerides predicted from Fink-FAT estimates. Two were known asteroids (426612 and 274218), whose positions were not compatible with the initial Fink-FAT trajectory. The remaining object was an irregular moon of Jupiter, Jupiter IX Sinope ($\approx$ 5.5 arcminutes from the Fink-FAT predictions). We found Sinope was also compatible with the initial Fink-FAT trajectory ($\leq$ 1 arcseconds distance from all alerts) and we concluded that \texttt{FF2023aaaaamb} was an observation of Sinope.

\paragraph{\texttt{FF2023aaaaalx}:} the last alert emission date for this trajectory was on 2022/08/22,  the observations were performed on 2022/09/25 from the OHP site and 2022/10/01 from the LCOGT site. The total arc is 12 days, and the orbital parameters were estimated to (a[AU], e, i[deg]) = (50.430875, 0.926643, 2.796635). In the OHP observations, there were two moving objects nearby the ephemerides predicted from Fink-FAT estimates. One was a known asteroid (426612), whose position was not compatible with the initial Fink-FAT trajectory. The remaining object was an irregular moon of Jupiter, Jupiter IX Sinope ($\approx$ 9 arcminutes from the Fink-FAT predictions). In the LCOGT observations, there were three moving objects nearby the ephemerides predicted from Fink-FAT estimates. Two were known asteroids (152295 and 425019), whose positions were not compatible with the initial Fink-FAT trajectory. The remaining object was an irregular moon of Jupiter, Jupiter IX Sinope ($\approx$ 10 arcminutes from the Fink-FAT predictions). We found Sinope was also compatible with the initial Fink-FAT trajectory ($\leq$ 1 arcseconds distance from all alerts), and concluded that \texttt{FF2023aaaaalx} was an observation of Sinope.

\paragraph{\texttt{FF2023aaaaamc}:} the last alert emission date for this trajectory was on 2022/08/29, and the observations were performed on 2022/10/01 from the LCOGT site. The total arc is eight days and the orbital parameters were estimated to (a[AU], e, i[deg]) = (2.358976, 0.251121, 5.275541). There was one moving object nearby the ephemerides predicted from Fink-FAT estimates, but it was a known asteroid (394919), whose position was not compatible with the initial Fink-FAT trajectory. {Hence, we have no confirmation for this object.}

\paragraph{\texttt{FF2023aaaaamd}:} the last alert emission date for this trajectory was on 2022/08/31, and the observations were performed on 2022/10/01 from the LCOGT site. The total arc is nine days, and the orbital parameters were estimated to (a[AU], e, i[deg]) = (6.525971, 0.783301, 4.540030). There were five moving objects nearby the ephemerides predicted from Fink-FAT estimates. There were all known asteroids (363563, 435953, 339694, 52703, 2015 BH451), whose positions were not compatible with the initial Fink-FAT trajectory. {Hence, we have no confirmation for this object.}

We note that during the processing of the observations at LCOGT of \texttt{FF2023aaaaalx}, four new moving objects previously unreported were also found (and not present in Fink as there were no ZTF observations at the same moment). These observations were sent to the Minor Planet Center.

\subsubsection{Including planet satellites}\label{sec:jupiter_moons}

We were not expecting to observe irregular satellites of Jupiter, but their ephemerides were not included in the MPC data files used by ZTF to associate alerts, so it is not surprising afterwards. Knowing this, we took all {65} unknown trajectories by Fink-FAT, and search for associations with Jupiter satellites compatible in terms of magnitude range (from JV Amalthea to JXX Taygete). We found seven trajectories associated with Sinope: four to Carme, three to Pasiphae, two to Ananke, one to Elara, and one to Himalia.

Knowing this, the orbital elements estimated by the default configuration of \orbfit\ are not correct, as these objects orbit around Jupiter. Not surprisingly, this is confirmed by Fig. \ref{fig:sso_futur} where all trajectories associated with Jupiter satellites have outliers values with respect to the rest of the trajectories where we mainly expect to recover main-belt asteroids with Fink-FAT. For completeness, we re-estimated the orbital elements from these observations but taking into account their relationship with Jupiter. As this functionality is not available in the publicly available \orbfit\ code source, we used the on-line \texttt{Find$\_$Orb} tool\footnote{\url{https://www.projectpluto.com/fo.htm}}. We provided the alert measurements in the PSV ADES format, and selected Jupiter as the element center to obtain the orbital elements. \review{The results are summarized in Table \ref{tab:extension_by_ephem_orbs}, where the rows correspond to \texttt{FF2023aaaaama} (1: Pasiphae), \texttt{FF2023aaaaamb} (2: Sinope), and \texttt{FF2023aaaaalx} (3: Sinope), respectively. Estimates are provided by the on-line \texttt{Find$\_$Orb} tool. The parameters are poorly constrained, as confirmed by the uncertainty parameter $U$ provided by the software for which values greater than nine denote an object's orbit extremely uncertain. One would need more observations to obtain more precise estimates.}

\begin{table}
\centering
\caption{Orbital parameters estimated from the three trajectories of the second follow-up campaign corresponding to Jupiter satellites, considering Jupiter as the center of mass.}
\begin{tabular}{lcccc}
 & a [AU] & e & i [deg] & U \\
\hline\hline 
1 & 0.233 $\pm$ 2.70 & 0.941 $\pm$ 0.235 & 97.7 $\pm$ 2.9 & 10.2 \\
2 & 0.103 $\pm$ 1.98 & 0.341 $\pm$ 0.198 & 153.2 $\pm$ 0.28 & 10.2 \\
3 & 0.168 $\pm$ 0.448 & 0.225 $\pm$ 4.18 & 159 $\pm$ 23 & 13.1 \\ 
\hline
\end{tabular}
\label{tab:extension_by_ephem_orbs}
\end{table}

\subsection{Limitations}\label{sec:limitations}

{In this section, we summarize the various limitations in the use of Fink-FAT that we identified over the course of this work:}

\begin{figure}[ht]
\includegraphics[width=9cm]{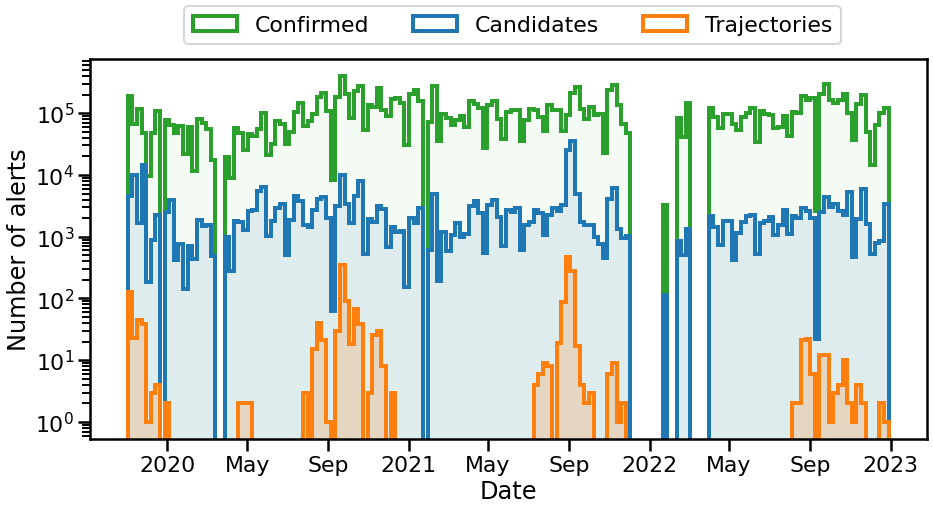}
\caption{Number of alerts from confirmed Solar System objects (green), Solar System candidates (blue), and alerts from trajectory candidates (orange) as a function of time. The bin width corresponds approximately to one week of data.}
\label{fig:season_effect}
\end{figure}

\begin{figure}[h]
\includegraphics[width=9cm]{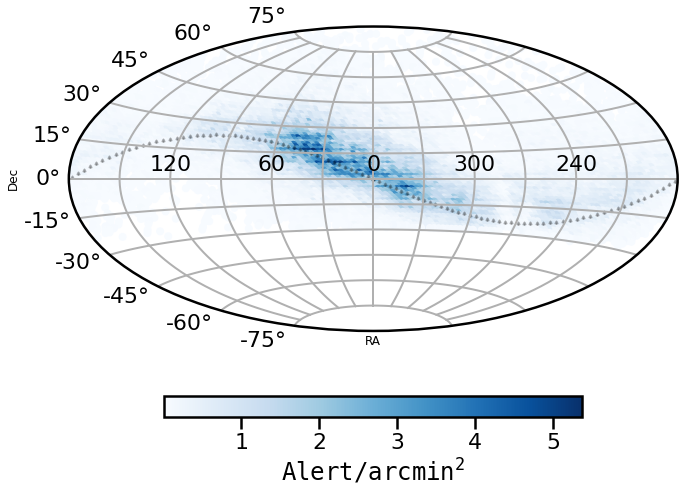}
\caption{{Footprint of the ZTF alert stream from November 2019 to December 2022 associated with confirmed Solar System objects (as in Fig. \ref{fig:skymap}), that also satisfy the detectability criterion (see Sect. \ref{sec:detectable}). We see an excess of alerts at (RA, Dec) = (0, 0), similarly to trajectory candidates. For reference, the ecliptic plane is shown with black triangles.}}
\label{fig:detectable_limitation}
\end{figure}

\begin{itemize}
\item Upon  receiving the alert, Fink refines the association with a potential confirmed Solar System object by relying only on distance criteria (see Sect. \ref{sec:sso_confirmed}). We plan to take into account in real time other association conditions, such as the co-linearity or magnitude difference using SkyBot.
\item Fink-FAT association steps (see Fig. \ref{fig:association_diagram}) are sequential. The associations found during a step are removed for the next step. Within a step, one trajectory can be extended with multiple measurements, but a measurement is only associated with one trajectory, and the association are also sequential. As a result, spurious associations can take over valid ones, \review{which drastically lowers the efficiency of Fink-FAT.} The inaccuracy of the association algorithm mainly drives this limitation. \review{Using an algorithm that improves the association accuracy such as the Kalman filter is a solution \citep{kalman} that we are presently investigating.}
\item  Fink-FAT parameters to search for new objects are based on the entire population of confirmed Solar System objects, without distinctions between dynamical classes (see Sect. \ref{sec:sso_confirmed}). As a result, this study is mainly driven by the population of main-belt asteroids detected by ZTF which are the most numerous. As we collect more objects over time, we plan to tune Fink-FAT for the search of other classes.
\item As we were not initially expecting to find alerts related to planet satellites, the orbit fitting step assumes an heliocentric system (see Sect. \ref{sec:jupiter_moons}). While the orbital solutions are somehow valid over a short period of time (we could retrieve the objects based on the predictions), we plan to systematically check for these in the future.
\item One of the limitation of Fink-FAT is the size of initial trajectories in terms of time and number of observations. Fink-FAT returns trajectories with a small number of points to limit the combinatorial, but also to quickly enable follow-up observations, but it does not try to aggregate more data in the future and refine the orbital parameters when possible. In our experiments with candidate Solar System objects, the largest trajectories had only nine observations and the smallest had six observations. The time between these observations is also very short (about nine days), and on average, the time between two subsequent observations was only two days. Due to these limitations, the orbits computed from these trajectories are often inaccurate, enabling an efficient follow-up only for a limited period of time. An extension of Fink-FAT is being considered to keep aggregating more data in the future and refine the initial orbital parameters as more data is processed. 
\item We found that the detection of the trajectories is not uniformly distributed over a single year, and most alerts from trajectory candidates are emitted in the period between August and December, as shown in see Fig. \ref{fig:season_effect}. First the ecliptic plane is higher in the sky from the ZTF observing site at this period \review{(higher in the sky so longer visibility, and observations with lower air mass)}. Second, due to weather condition at the observing site, the period of January to March is less suitable for observations (see, e.g., the alert coverage\footnote{\url{https://fink-portal.org/stats}}). Third, there were long maintenance \review{periods} of the ZTF camera during December and April of 2022, reducing the number of observations. We also suspect a correlation with the method, but we cannot firmly conclude at this stage, as this pattern is not as strong in the confirmed objects nor in the Solar System candidates (there is some oscillation, but the range between extrema in the number of alerts selected is smaller). We are still investigating.
\item We found that most of the trajectory candidates are concentrated around (RA, Dec) = (0, 0) in the sky (see Fig. \ref{fig:skymap}). This is typically linked to the seasonal variations mentioned above, but we also found a correlation with our method to select valid alerts to form trajectories. For example, we took all alerts associated with confirmed Solar System objects, we kept only those satisfying the criterion of detectability (as defined in Sect. \ref{sec:detectable}), and we project these alerts on the sky. The results are shown in Fig. \ref{fig:detectable_limitation}, where we clearly see an excess of alerts around (RA, Dec) = (0, 0). It is not clear whether the cadence of the survey also plays a role here and we are still investigating this aspect.
\end{itemize}

\section{Conclusion and perspective for LSST}

{The use of an alert broker to overcome the challenges posed by the linkage problems in the context of large volumes of alert data, by reducing the initial number of inputs to link, has proven useful for the real-time identification of Solar System object. Based on this approach, we developed a new component in Fink, Fink-FAT, to detect potential new asteroids. Fink-FAT works in two steps: the association step which relies on a linking algorithm using simple dynamical consideration (co-linearity, magnitude evolution, and apparent motion) and the orbit fitting step which relies on the \orbfit\ software. }

\review{Fink-FAT has been successfully applied on the Solar System alert data stream produced by Fink from the ZTF alert data stream. The parameters of the algorithm were tuned using confirmed Solar System objects in the ZTF alert stream, and applied to Solar System candidate alerts selected by Fink. The low efficiency (25-45$\%$) of Fink-FAT remains its main bottleneck. This is due to the fact that we  sequentially apply the association steps, discarding the associated elements from the possible association for the next steps. On the other hand, the purity of the algorithm reaches 97$\%$ after full orbit estimates, which is a requirement for performing efficient follow-up observations.}

Fink-FAT has been also tested for LSST-like alert stream, and it demonstrated that it is particularly well adapted in the context of large alert data streams \review{for Solar System candidates}: it requires modest hardware resources to operate, while having a relatively low computational time. We note though that if Fink-FAT is less prone to cadence effect than MOPS for example (as it does not only rely on tracklets), it is not as cadence-independent as other recent more sophisticated association algorithms might be, such as THOR \citep{moeyens2021thor}.

{The two follow-up campaigns enabled to test some aspects of Fink-FAT operations. Despite the rather large delay between the initial trajectories and the follow-up observations (more than a month), four trajectories out of six turned out to be associated with real objects from the Solar System based on Fink-FAT predictions on small arcs. The distances of the objects to their predictions were within the expectations shown in Fig. \ref{fig:deviation_follow_up}. For the two remaining trajectories, we can speculate that if they were initially associated with real moving objects, the deviation of the prediction from the true position would have been beyond the field of view of the telescope (27 arcminutes for the LCOGT). Overall, even if no new Solar System object was reported from Fink-FAT trajectories for these two observation campaigns, it confirms the ability of Fink-FAT to form coherent trajectories.}

{Fink-FAT is deployed as a real-time component in Fink since 2022. Each night, the system creates or extends the pool of trajectories and fits orbits for those that exceed a certain number of points. Finally, the Solar System candidate alerts, the trajectories, and their orbital parameters are entered into the Fink database. All outputs are publicly available via the different interoperable services of Fink\footnote{\url{https://fink-broker.readthedocs.io}}. In addition, a new area in the Fink Science Portal is being developed to allow users to perform further analyses directly in their browser and easily plan follow-up observations.}

\begin{acknowledgements}
\review{RLM and JP thank the anonymous referee for invaluable comments on the initial version of the paper.} RLM and JP thank Emille O. Ishida, Anais M\"oller and Nicolas Leroy for their useful feedback on the initial draft. This work was developed within the Fink community and made use of the Fink community broker resources. Fink is supported by LSST-France and CNRS/IN2P3. This project has received financial support from the CNRS through the MITI interdisciplinary programs and from CNES.
\end{acknowledgements}

\bibliographystyle{aa}
\bibliography{aanda}

\begin{appendix}

\section{Fink-FAT pseudo-code} \label{app:pseudo_code}

\begin{algorithm}
    \caption{Intra-night association algorithm}\label{code:intra}
    \begin{algorithmic}[1]
    \Function{intra\_night\_association}{$A_i$}
    \State $T_{intra} \gets \emptyset$
    \State $T_{intra} \gets \{(a_j, a_k) | \forall a_j, a_k \in A_i, \delta d < sep\_limit\}$ \Comment{using a KD-tree}
    \State $T_{intra} \gets T_{intra} \setminus \{(a_j, a_k) | \neg cut\_2\}$
    \State $T_{intra} \gets T_{intra} \setminus \{(a_j, a_k) | \neg cut\_3\}$
    \ForEach {$t_0=(a_0, a_1), t_1=(b_0, b_1) \in T_{intra} $}
    \If {$a_1 R_{intra} b_0$}
    \State $T_{intra} \gets T_{intra} \setminus {t_0, t_1}$
    \State $t \gets (a_0, a_1, b_0, b_1)$
    \State $T_{intra} \gets T_{intra} \cup \{t\}$
    \EndIf
    \EndFor
    \State \textbf{return} $T_{intra}$
    \EndFunction
    \end{algorithmic}
\end{algorithm}

\begin{algorithm}
    \caption{Fink-FAT algorithm}\label{code:finkfat}
    \begin{algorithmic}[1]
    \Function{Fink-FAT}{$T, O, A_i$}
    \State $T_{inter} \gets \emptyset$
    \State $T_{intra} \gets intra\_night\_association(A_i)$
    \State $A_i \gets A_i \setminus \{ a_j | \forall a_j \in T_{intra} \}$
    \ForEach {$t=(a_0, ..., a_k) \in T $}
    \ForEach{$t'=(b_0, ..., b_k) \in T_{intra} $}
    \If {$ P(a_{k-1}, a_k, b_0) $}
    \State $T \gets T \setminus \{t\}$
    \State $T_{intra} \gets T_{intra} \setminus \{t'\}$
    \State $t_{merge} \gets (a_0, ..., a_k, b_0, ..., b_k)$
    \State $T \gets T \cup \{ t_{merge} \}$
    \EndIf
    \EndFor
    \ForEach{$b \in A_i $}
    \If {$P(a_{k-1}, a_k, b_0)$}
    \State $T \gets T \setminus \{t\}$
    \State $A_i \gets A_i \setminus \{b\}$
    \State $t_{merge} \gets (a_0, ..., a_k, b)$
    \State $T \gets T \cup \{ t_{merge} \}$
    \EndIf
    \EndFor
    \EndFor
    \ForEach{$t=(b_0, ..., b_k) \in T_{intra} $}
    \ForEach{$o \in O$}
    \If {$P(o, b_0, b_1)$}
    \State $T_{intra} \gets T_{intra} \setminus \{t\}$
    \State $O \gets O \setminus \{o\}$
    \State $t_{merge} \gets (o, b_0, ..., b_k)$
    \State $T \gets T \cup \{ t_{merge} \}$
    \EndIf
    \EndFor
    \EndFor
    \State $T_{new} \gets \{(a_j, a_k) | a_j R_{inter} a_k, \forall a_j \in O, \forall a_k \in A_i\}$
    \State $O \gets O \setminus \{a_j | \forall a_j \in T_{new}\}$
    \State $A_i \gets A_i \setminus \{a_k | \forall a_k \in T_{new}\}$
    \State $T \gets T \cup T_{intra} \cup T_{new}$
    \State $O \gets O \cup A_i$
    \State \textbf{return} $T, O$
    \EndFunction
    \end{algorithmic}
\end{algorithm}

\section{Extending parameters used in Fink-FAT} \label{app:nea}

As described in Sect. \ref{sec:sso_confirmed}, the parameters used in Fink-FAT are derived from the alerts returned by ZTF with a MPC match without taking into account the orbit types. Hence, the values of the parameters are mainly driven by the population of main belt asteroids detected by ZTF, which are the most numerous. To better probe the impact of such a choice in the recovery of objects in different groups, we re-estimated Fink-FAT parameters but based only on objects from the near-Earth asteroid group (see, e.g., Table \ref{tab:sso_condition_inter_nea}, derived from 1,970 objects in the confirmed SSO dataset between 2019 and 2023). As the Fink-FAT parameters are set from their cumulative distribution, we effectively extend the targeted group to NEA, but main-belt objects are still included (as they typically evolve slowly). 

The total Fink-FAT runtime increased significantly compared to the case with the default set of parameters (user time of 15 hours, using the cluster mode for the orbit fitting step \review{with the same hardware configuration than in Sect. \ref{section_results}}). This increase of time is due to the higher number of associations formed and trajectories to fit, allowed by the extended Fink-FAT input parameters. Conversely, there are fewer trajectories with an orbit estimate (213 compared to 327) for a total of 1,316 linked observations. The decrease of the number of trajectories is due to a higher false positive rate when associating alerts: Fink-FAT produces many trajectories intersecting, which are then discarded (\review{see discussion on the efficiency in Sect. \ref{sec:comparison}}). We note though that the trajectory with the smaller arc length reaches 0.1 day (six alerts in the same night).

\begin{figure*}[h]
\includegraphics[width=\textwidth]{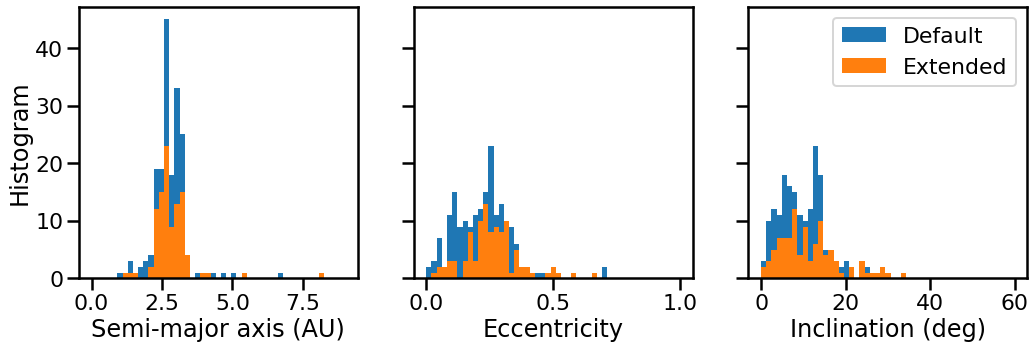}
\caption{Distribution of orbital parameters from reconstructed trajectories with full orbit estimate using the default Fink-FAT parameters (blue histograms; see Table \ref{tab:sso_condition_inter}), and the extended set of parameters estimated from the NEA group (orange histograms; see Table \ref{tab:sso_condition_inter_nea}). The left panel shows the distribution of the semi-major axis parameter, the middle panel shows the distribution of the eccentricity parameter, and the right panel shows the distribution of the inclination parameter.}
\label{fig:comparison_default_nea}
\end{figure*}

The orbital parameter distributions are however similar to the distribution of parameters estimated from the default case described in Sect. \ref{section_discovery} and shown in Fig. \ref{fig:comparison_default_nea}; especially as  there is no excess of objects with a small semi-major axis in the extended case. Our interpretation is that even if the Fink-FAT parameter space has been extended, the results are still driven by the main bulk of objects from the main belt, and we would need to include more objects from the NEA group when estimating Fink-FAT parameters to efficiently reconstruct similar trajectories.

\begin{table}
\centering
\caption{Same as Table \ref{tab:sso_condition_inter}, but using only the objects matched to near-Earth asteroids from the confirmed SSO dataset.}
\begin{tabular}{ll}
\hline\hline
Inter-night & \\
$r_d$ &  1.0 deg/day \\
$r_m$ (from same filter bands) & 0.1 mag/day  \\
$r_m$ (from different filter bands) & 0.8 mag/day  \\
$r_\alpha$ & 0.6 deg/day \\
\hline\hline
Intra-night & \\
$r_d$ &  0.05 deg \\
$r_m$ (from same filter bands) & 0.2 mag  \\
$r_m$ (from different filter bands) & 0.6 mag  \\
$r_\alpha$ & -- \\
\hline
\end{tabular}
\label{tab:sso_condition_inter_nea}
\end{table}

\section{Cataloguing and data availability} \label{sec:api}

\review{All new Solar System object candidates found by Fink-FAT are stored permanently in the Fink database, and publicly available as soon as the observing night is finished. We deployed an endpoint in the Fink REST API that lets users query this information\footnote{See \url{https://fink-portal.org/api} for more information}. Users can retrieve the photometry of objects related to candidate orbits (\texttt{kind=lightcurves}) and the orbital parameters for orbit candidates (\texttt{kind=orbParams}). We propose an example of a query to retrieve all the orbital parameters, formulated in the Python programming language, as follows:}

\begin{lstlisting}[language=Python, frame=single, caption={}, label=listing:tracklet]
import requests

r = requests.post(
  'https://fink-portal.org/api/v1/ssocand',
  json={
    'kind': 'orbParams'
  }
)
\end{lstlisting}

\end{appendix}
 
\end{document}